\documentclass[usenatbib]{mnras}
\usepackage{makecell}  
\usepackage{lmodern}             
\usepackage{microtype}        
\usepackage{booktabs}
\usepackage{xcolor}
\usepackage{tabularx}
\usepackage{multirow} 
\usepackage{newtxtext,newtxmath,amsmath,upgreek}
\usepackage[percent]{overpic}
\usepackage{tikz}
\usepackage{caption}
\usepackage{ragged2e}            
\usepackage{subcaption}
\numberwithin{equation}{section}
\usepackage[percent]{overpic}
\usepackage[T1]{fontenc}

\DeclareRobustCommand{\VAN}[3]{#2}
\let\VANthebibliography\thebibliography
\def\thebibliography{\DeclareRobustCommand{\VAN}[3]{##3}\VANthebibliography}

\usepackage{graphicx}	
\usepackage{amsmath}	
\usepackage{siunitx} 

\newcommand{\CCC}{{\it Cloud-Cloud Collision}}
\newcommand{\CCCs}{{\it Cloud-Cloud Collisions}}
\newcommand{\ACC}{{\it Layer Accumulation}}
\newcommand{\acc}{_{\mbox{\tiny ACC}}}
\newcommand{\LAT}{{\it Lateral Contraction}}
\newcommand{\lat}{_{\mbox{\tiny LAT}}}
\newcommand{\FRAG}{{\it Layer Fragmentation}}

\newcommand{\SPW}{{\it Spiders-Web}}
\newcommand{\SPWS}{{\it Spiders-Web System}}
\newcommand{\HF}{{\it Hub-Filament}}
\newcommand{\HFS}{{\it Hub-Filament System}}
\newcommand{\HFSs}{{\it Hub-Filament Systems}}
\newcommand{\thSB}{\theta_{\mbox{\tiny SB}}}
\newcommand{\Msun}{{\rm M}_{_\odot}}

\title[The influence of magnetic fields in Cloud-Cloud Collisions]{The influence of magnetic fields in Cloud-Cloud Collisions.}
\author[T. Georgatos \& A.P. Whitworth]{
Theotokis Georgatos,\thanks{E-mail: GeorgatosT@Cardiff.AC.UK}
Anthony P. Whitworth,
\\
School of Physics and Astronomy, Cardiff University, Queens Buildings, The Parade,Cardiff CF24 3AA, Wales, UK\\
}

\date{Accepted XXX. Received YYY; in original form ZZZ}

\pubyear{XXXX}

\begin{document}
\label{firstpage}
\pagerange{\pageref{firstpage}--\pageref{lastpage}}
\maketitle

\parskip=0.0cm
\begin{abstract}
Cloud-cloud collisions are expected to trigger star formation by compressing gas into dense, gravitationally unstable regions. However, the role of magnetic fields in this process is unclear. We use SPH to model head-on collisions between two uniform density clouds, each with mass $500\,\Msun$, initial radius 2 pc, and threaded by a uniform magnetic field parallel to the collision velocity. As in the non-magnetic case, the resulting shock-compressed layer fragments into a network of filaments. If the collision is sufficiently slow, the filaments are dragged into radial orientations by non-homologous gravitational contraction, resulting in a \HF\ morphology, which spawns a centrally concentrated monolithic cluster with a broad mass function shaped by competitive accretion and dynamical ejections. If the collision is faster, a \SPW\ of intersecting filaments forms, and star-systems condense out in small sub-clusters, often at the filament intersections; ~due to their smaller mass reservoirs, and the lower probability of dynamical ejection, the mass function of star-systems formed in these sub-clusters is relatively narrow. Magnetic fields affect this dichotomy quantitatively by delaying collapse and fragmentation. As a result, the velocity threshold separating \HF\ and \SPW\ morphologies is shifted upward in magnetised runs, thereby enlarging the parameter space in which \HFSs\ form, and enhancing the likelihood of producing centrally concentrated clusters. Thus magnetic fields regulate both the morphology and timing of star formation in cloud-cloud collisions: they broaden filaments, delay the onset of star formation, and promote the formation of \HFSs\, monolithic clusters and high-mass star-systems.
\end{abstract}


\section{Introduction}\label{SEC:Intro}

\CCCs\ are a compelling mechanism for triggering star formation \citep{HabeAOhtak1992PASJ44p203, InoueTFukuiY2013ApJ774L31, WuBetal2015ApJ811A56, WuBetal2017aApJ835A137, WuBetal2017bApJ841A88, FukuiYetal2021aPASJ73S1, HorieSetal2024MN527p10077, MaityAetal2024ApJ974A229, WeisMetal2024MN532p1262}. When two clouds collide at supersonic speed, they produce a dense, shock-compressed layer, which rapidly accumulates mass and may become gravitationally unstable. A significant role is played by the ram pressure of continuing gas inflow, making the shock-compressed layer a conducive environment for the formation of prestellar cores, especially massive cores \citep[e.g.][]{HsuCJetal2023MN522p700}. It is therefore important to explore what effect a magnetic field has on this scenario.
 
The frequency  of \CCCs\ —and consequently their cumulative impact on star formation—is difficult to quantify \citep[see][]{HaworthTetal2015aMN450p10, GongYetal2019AA632A115}. On the basis of hydrodynamic simulations, \citet{DobbsCetal2015MN446p3608} conclude that clouds experience very few collisions during their lifetime, and therefore cloud evolution is little affected by collisions. On the other hand, \citet{BalfourSetal2017MN465p3483} estimate that relatively low-mass, low-velocity \CCCs\ are frequent in the Galaxy, and could play a dominant role in regulating the overall rate of star formation. This is partly down to what is understood by a \CCC\ (see Section \ref{SEC:GenConsid})

Observationally, identifying cases of star formation triggered by \CCCs\ remains challenging, thereby limiting our ability to assess the influence of \CCCs\ on star formation. Certain diagnostic features, for example a "bridge" feature connecting two colliding clouds in position-velocity space, have been proposed as indicators \citep[e.g.][]{HaworthTetal2015aMN450p10, FukuiYetal2018aPASJ70S44, FukuiYetal2018bPASJ70S46, MaityAetal2024ApJ974A229}. However, these signatures are often difficult to detect unambiguously, due to their strong dependence on viewing angle and chemistry \citep[e.g.][]{PriestleyFWhitworthA2021MN506p775}.

There have been several studies of star formation triggered by \CCCs\ \citep[for example][]{WhitworthAPetal1994aAA290p421, WhitworthAPetal1994bMN268p291, KitsionasSWhitworthA2007MN378p507, WhitworthAP2016MN458p1815, DinnbierFetal2017MN466p4423, WhitworthAPetal2018PASJ70S55, PriestleyFWhitworthA2021MN506p775, WhitworthAPetal2022MN517p4940}. ~\citet{BalfourSetal2015MN453p2471} use pure hydrodynamic simulations to demonstrate that a low-velocity collision produces a shock compressed layer, which then fragments into a network of filaments. They show that the shock-compressed layer fragments into two distinct morphologies depending on whether the initial collision velocity exceeds a velocity threshold \(\Delta u_{\rm thresh}\).

For high-speed collisions (relative velocity greater than $\Delta u_{\rm thresh}\sim 3.0\,{\rm km\,s^{-1}}$) the layer builds up sufficiently fast that fragmentation occurs before the layer has had time to contract laterally (i.e. orthogonal to the collision axis). This results in a network of intersecting filaments resembling a \SPW.  Low-mass cores form out of the filaments, often at their intersections, and a small sub-cluster condenses out of each core. The mass reservoir in each core is low, so it is hard for very massive stars to form there. The number of stars is small, so there is little opportunity to dynamically eject low-mass stars, thereby terminating their growth. As a result the stellar mass function is relatively narrow. Only when star formation is largely complete do these sub-clusters fall together and relax to form a larger cluster, so the mass function remains relatively narrow.

In contrast, for collisions with relative velocities below $\Delta u_{\rm thesh}$, the layer builds up more slowly and at the same time contracts laterally and non-homologously (i.e. faster towards the centre). This non-homologous contraction drags the forming filaments into radial orientations so that by the time they start to fragment they are already feeding their material towards a monolithic central massive core, resulting in an \HF\ morphology. In the central massive core, stars grow by competitive accretion. Consequently a few stars attain very high masses, and others are ejected dynamically thereby cutting off accretion and limiting them to very low masses. The result is a centrally concentrated monolithic cluster with a broad mass function.

A critical issue is then whether magnetic fields influence the separatrix between circumstances that produce \SPW\ morphologies and those that produce \HF\ morphologies. Since the underlying filamentary morphology governs how material is delivered to protostars, any such magnetic regulation is expected to have a direct impact on the spatial distribution of newly-formed stars and their mass function.
 
MHD simulations of high velocity \CCCs\ by \citet{InoueTFukuiY2013ApJ774L31} and \citet{FukuiYetal2021bPASJ73S405} demonstrate that collisions between clouds of $\sim 10^4\,\textrm{M}_\odot$ with relative velocities of $10\,{\rm km\,s^{-1}}$ can produce massive fragments of order $50\,\textrm{\,M}_\odot$. This suggests that massive stars may form in high-velocity collision environments.

Simulations of cloud–cloud collisions between unequal-mass clouds by \citet{SakreNetal2021PASJ73S385}, and spanning a range of magnetic field strengths and collision velocities up to $10\,\mathrm{km\,s^{-1}}$, suggest that stronger magnetic fields can facilitate the formation of high-mass cores. They attribute this conclusion to the suppression of low-mass core growth and the stabilisation of the shock-compressed layer.

In contrast, on the basis of simulations involving higher collision velocities in the range 20 to $40\,\mathrm{km\,s^{-1}}$, \citet{SakreNetal2021PASJ73S385} conclude that magnetic fields may inhibit massive core formation by limiting post-collision gas accretion. Enhanced magnetic pressure drives rapid expansion of the shock-compressed region, leading to the disruption of gravitationally unbound cores. This implies that collision velocity is a key factor in determining whether magnetic fields promote or suppress massive core growth.
 
These MHD studies of \CCCs\ demonstrate that magnetic fields alter fragmentation and core formation in colliding clouds \citep[e.g.][]{InoueTFukuiY2013ApJ774L31, FukuiYetal2021bPASJ73S405, SakreNetal2021PASJ73S385, SakreNetal2023MN522.p4972}, but we are unaware of any studies that have systematically explored how varying both the initial field strength, and the collision velocity, affects filament widths, stellar mass functions, and measures of stellar clustering.

In this study, we model \CCCs\ using initial conditions similar to those of \citet{BalfourSetal2015MN453p2471}, but with the addition of a magnetic field aligned parallel to the collision axis. Our aim is to identify, and where possible quantify, the effects of a magnetic field on the resulting star formation, using very simplistic initial conditions and constitutive physics, rather than to reproduce accurately all the features of observed systems. Therefore we only vary two model parameters (the collision velocity and the initial magnetic field strength), and we refer to our models as numerical experiments, rather than simulations.

The remainder of the paper is organised as follows. Section \ref{Sec:NumercialMethods} describes the initial conditions for the numerical experiments. Section \ref{SEC:PhysProc} sketches the main physical processes at work in the experiments. Section \ref{Sec:Results} presents the results. In Section \ref{SEC:Discussion}, we discuss the implications of the results, and in Section \ref{SEC:Conclusions} we summarise our main conclusions.

\section{Numerical Methods and Initial Conditions}\label{Sec:NumercialMethods}

\begin{table}
\caption{Parameters describing the initial conditions.}
\begin{center}
\begin{tabular}{ll}\hline
Initial cloud radius & $R_{\rm o} = 2$\,pc \\
Cloud mass & $M_{\rm o} = 500\,\Msun$ \\
Initial cloud density & $\rho_{\rm o} \simeq 10^{-21}$\,g\,cm$^{-3}$ \\
Isothermal sound speed in cloud& $a_{\rm {o}}$ $= 0.187$\, km\,s$^{-1}$ \\
Background density & $\rho_{\rm back} = 10^{-23}$\,g\,cm$^{-3}$ \\
Isothermal sound speed in background$\;\;$ & $a_{\rm back} \simeq 1.87$\, km\,s$^{-1}$ \\
Relative collision velocities & $\Delta u_{\rm o}=2u_{\rm o} = 2.4,\,2.8,$ \\
 & \hspace{0.15cm} $3.2,\,3.6,\,4.0,\,4.4\,{\rm km\,s^{-1}}$ \\
 Mach Number of collision & ${\cal M}_{\rm o}=u_{\rm o}/a_{\rm o}=6.4,\,7.5,$ \\
 & \hspace{0.4cm} $8.5,\,9.6,\,10.7,\,11.7$ \\
Magnetic fields  & $B_{\rm o} = 0,\,1,\,2,\,3.3,\,5\,\upmu$G \\
Sink creation density threshold & $\rho _{\rm sink} = 10^{-16}$\,g\,cm$^{-3}$ \\
Sink radius & $R_{\rm sink} = 600\,{\rm au} = 0.003\,{\rm pc}$ \\
Size of the computational domain & $L_{\rm o} = 8$\,pc \\
Number of SPH particles & $\mathcal{N}_{\rm SPH} \simeq 10^6$ \\\hline
\end{tabular}
\end{center}
\label{tab:glossary}
\end{table}

\begin{table}
\caption{Glossary}
\begin{center}
\begin{tabular}{ll}\hline
\CCC\ & a cloud-cloud collision involving two \\
 & clouds with initial velocities $u_{_x}\!=\!\pm u_{\rm o}$ \\
\ACC\ & accumulation of a shock-compressed layer \\
 & following a \CCC\ \\
\LAT\ & reduction of the lateral extent of this \\
 & shock-compressed layer \\
\FRAG\ & break-up of the shock-compressed layer, \\
 & driven by self-gravity\\
\SPWS\ & an isotropic network of randomly oriented \\
 & filaments (e.g. bottom left panel of Fig. \ref{fig:ColDen}) \\
\HFS\ & an isotropic array or radially oriented \\
 & filaments (e.g. top right panel of Fig. \ref{fig:ColDen}) \\\hline
\end{tabular}
\end{center}
\end{table}

We model \CCCs\ using the publicly available Smoothed-Particle Magneto-Hydrodynamics (SPH) code \textsc{Phantom} \citep{PriceDetal2018PASA35e031}, with initial conditions analogous to those in \citet{BalfourSetal2015MN453p2471}.  The default \citet{morris1997switch} viscosity prescription is adopted.

Each cloud has mass $M_{\rm o}=500\,\Msun$ and radius $R_{\rm o}=2$\,pc, corresponding to a uniform density $\rho_{\rm o}= 10^{-21}$\,g\,cm$^{-3}$. The cloud gas has isothermal sound speed $a_{\rm o}=0.187\,{\rm km\,s^{-1}}$, corresponding to molecular gas with solar elemental abundances and gas-kinetic temperature $T_{\rm o}=10\,{\rm K}$.   
  
The cloud centres are initially located at
\begin{eqnarray}
[x,y,z] & = & [\pm 2.0,0,0] \text{\,pc},
\end{eqnarray}
with bulk velocities
\begin{eqnarray}
[u_x,u_y,u_z] & = & [\mp u_{\rm o},0,0].
\end{eqnarray}
Consequently the clouds immediately collide head-on, with relative velocity $\Delta u_{\rm o}\!=\!2u_{\rm{o}}$.  

The colliding clouds are envisaged as part of a larger cloud complex with mass in the range $\sim\!2,000$ to $\sim\!20,000\,\Msun$. Thus, adopting Larson’s scaling relations for the velocity dispersion within a cloud complex \citep[][$\sigma_\upsilon\sim0.42\,{\rm km\,s^{-1}}(M/\Msun)^{0.20}$]{LarsonRB1981MN194p809}, we assign cloud velocities $u_\textrm{o}\sim 1.0$, $1.2$, $1.4$, $1.6$, $1.8$, $2.0$, $2.2$ and $2.4\,{\rm km\,s^{-1}}$. The relative velocities of the clouds are therefore $\Delta u_{\rm o}\sim 2.0$, $2.4$, $2.8$, $3.2$, $3.6$, $4.0$, $4.4$ and $4.8\,{\rm km\,s^{-1}}$.

The clouds are embedded in a hot, low-density background with density $\rho_{\rm back}=\rho_{\rm o}/100$ and isothermal sound speed $a_{\rm back}=10a_{\rm o}$. This ensures approximate pressure balance at the cloud boundaries.
 
The initial magnetic field strength, $B_{\rm o}$, is set to a fraction, $\beta_{\rm o}$, of the critical value for the magnetic field to balance the self-gravity of the cloud \citep{ChandrasekharSFermiE1953ApJ118p116},
\begin{eqnarray}\label{EQN:Bcrit.01}
B_{\rm crit} & = & 2\pi G^{1/2}\rho_{\rm o}R_{\rm o}, \\\label{EQN:Bo.01}
B_{\rm o} & = & \beta_{\rm o} B_{\rm crit}.
\end{eqnarray}
We adopt five magnetic field strengths corresponding to $\beta_{\rm o} = 0, 1/10, 1/5, 1/3, \text{ and } 1/2$. This yields field strengths of $B_{\rm o} \sim 0, 1, 2, 3.3 \text{ and } 5 \, \upmu$G. We have checked that -- for the collision velocities considered here -- fragmentation of the shock-compressed layer into filaments is entirely suppressed with $\beta_{\rm o} = 1\;(B_{\rm o}\!=\!10\,\upmu{\rm G})$. The uniform initial magnetic field is parallel to the collision axis $x$, i.e. ${\boldsymbol B}=[B_{\rm o},0,0]$. This uniform magnetic field threads the entire computational domain.

The computational domain is a cube with side $L_{\rm o}=8$\,pc and we adopt periodic boundary conditions. The initial distribution of SPH particles is generated by relaxing a periodic box into a glass-like distribution and then cutting out a sphere with radius $R_{\rm o}=2\,{\rm pc}$ to represent a cloud. The sphere is then rotated through random Euler angles. Different random rotations are used for the two clouds. To generate the background a similar glass distribution is generated, with 100 times lower particle density, and two spherical holes are cut to accommodate the clouds. The SPH particles representing the background are not rotated. The clouds are modelled with $\,\mathcal{N}_{\rm SPH} \simeq 10^6$ particles and each particle has approximately $57$ neighbours, yielding a minimum resolved mass of $\sim\!0.1\,\Msun$.

Sink particles are introduced at a density threshold $\rho_{\rm sink} = 10^{-16}\,{\rm g\,cm^{-3}}$, using the \citet{BateMetal1995MN277p362} prescription. Sink merging is turned off, and the sink radius is set to $R_{\rm sink}=600\,{\rm au}$, with only gas within $180\,{\rm au}$ being unconditionally accreted.

We refer to sinks as star-systems, because our numerical experiments are unable to resolve either below the Hydrogen Burning Limit, or any but the widest binary orbits. Thus `star-systems' could be single stars, binaries and higher-order multiples, or even just small sub-clusters that have not had time to relax dynamically.

The parameters of the initial conditions and constitutive physics are summarised in Table \ref{tab:glossary}. For each combination of $B_{\rm o}$ and $\Delta u_{\rm o}$ we run three realisations, each with different initial distributions of SPH particles (i.e. different random rotation angles). 

Since the model does not include stellar feedback, we terminate experiments once $10 \%$ of the mass has been converted into star-systems, since by then feedback processes are likely to be important. We refer to this time as $t_{10\%}$, and most of the statistics that we present are collated at $t_{10\%}$, combining the three realisations of each $(B_{\rm o},\Delta u_{\rm o})$ combination.

In the sequel, the $x$ axis -- which is parallel or anti-parallel to both the initial cloud velocities and the initial magnetic field -- is described as the collision axis; vector components parallel to the $x$ axis are described as `axial' and denoted with subscript `$x$', ~e.g. axial magnetic field $B_x$. Conversely, vector components orthogonal to the $x$ axis are termed `lateral', and denoted with a subscript `$yz$', ~e.g. lateral magnetic field $B_{yz}\equiv(B_y^2+B_z^2)^{1/2}$.

\section{Global Physical processes}\label{SEC:PhysProc}

As in the non-magnetic case \citep{BalfourSetal2015MN453p2471}, the colliding clouds produce a shock-compressed layer, which may eventually become sufficiently massive to fragment into filaments and cores, and spawn star-systems. In this section we isolate and describe the critical global physical processes which contribute to the evolution of colliding clouds.

\subsection{No magnetic field, fast collision}

The simplest setup involves no magnetic field, $B_{\rm o}\!=\!0$, and a relatively high collision speed, say $\Delta u_{\rm o}\gtrsim 3.2{\rm km\,s^{-1}}$ (${\cal M}_{\rm o}\gtrsim 17$). There are then just two basic processes, \ACC\ and \FRAG, and they are approximately sequential. 

\ACC\ describes the assembly of the layer by the colliding clouds. The surface-density builds up to $\Sigma_{\mbox{\tiny MAX}}\!\sim\!4R_{\rm o}\rho_{\rm o}$ at a rate $\dot\Sigma\!\sim\!\rho_{\rm o}\Delta u_{\rm o}$, and therefore on a timescale $t\acc\!\sim\!4R_{\rm o}/\Delta u_{\rm o}$.

\FRAG\ describes the break-up of this layer due to self-gravity \citep{MiyamaSetal1987aPThPh78p1051, MiyamaSetal1987bPThPh78p1273}, and requires $\Sigma_{\mbox{\tiny MAX}}\!\gtrsim\!a_{\rm o}^2/2\pi GR_{\rm o}$, or equivalently $M_{\rm o}/R_{\rm o}\!\gtrsim\!a_{\rm o}^2/4G$. Fragmentation proceeds through the formation of a network of intersecting filaments, which we term a \SPW. A multitude of small cores then forms, often at the intersections of the filaments, and then star-systems condense out of these cores.

The individual cores have a small mass reservoir, and consequently the stars formed in them have a relatively small mass-range. There is insufficient mass to form very high-mass stars. There are also too few stars for many very low-mass stars to be formed by having their ongoing accretion abruptly terminated by dynamical ejection from the core.

The star-systems formed from the different individual small cores eventually fall together and relax to form a single star cluster, but by then the masses of the stars have essentially been determined, and further growth by accretion is a small perturbation to the overall mass spectrum.

\subsection{No magnetic field, slow collision}

If there is still no magnetic field, $B_{\rm o}\!=\!0$, but the collision is relatively slow, say $\Delta u_{\rm o}\lesssim 2.4\,{\rm km\,s^{-1}}$ (${\cal M}_{\rm o}\lesssim13$), a third process comes into the reckoning, namely \LAT. 

\LAT\ describes the convergence of material towards the $x$ axis, driven by the global self-gravity of the two clouds, and it occurs on a timescale $t\lat\!\simeq\!\left\{3G\rho_{\rm o}\right\}^{-1/2}$. This has two important effects.

Firstly, during \ACC, \LAT\ causes the approximately circular patch where the clouds collide to shrink, and therefore the build-up of the critical surface-density is accelerated.

Secondly, during \FRAG, \LAT\ is non-homologous: the central parts of the layer converge on the $x$ axis faster than the outer parts. This drags the forming filaments into radial orientations, so that instead of feeding material into a multitude of small isolated cores, the filaments channel material into a central monolithic hub. Here, competitive accretion promotes the formation of very high-mass stars. In addition, very low-mass stars are created more frequently by having their on-going accretion abruptly terminated by dynamical ejection.

Some of the material arriving in the central core has already been converted into stars, but much of it arrives in the form of gas, and is therefore available, either to be accreted by existing stars, or to form additional stars. Consequently the stars form and grow in a single proto-cluster, and the resulting star cluster has a broad mass spectrum.

\begin{figure*}
\centering
\begin{overpic}[width=1.03\linewidth]{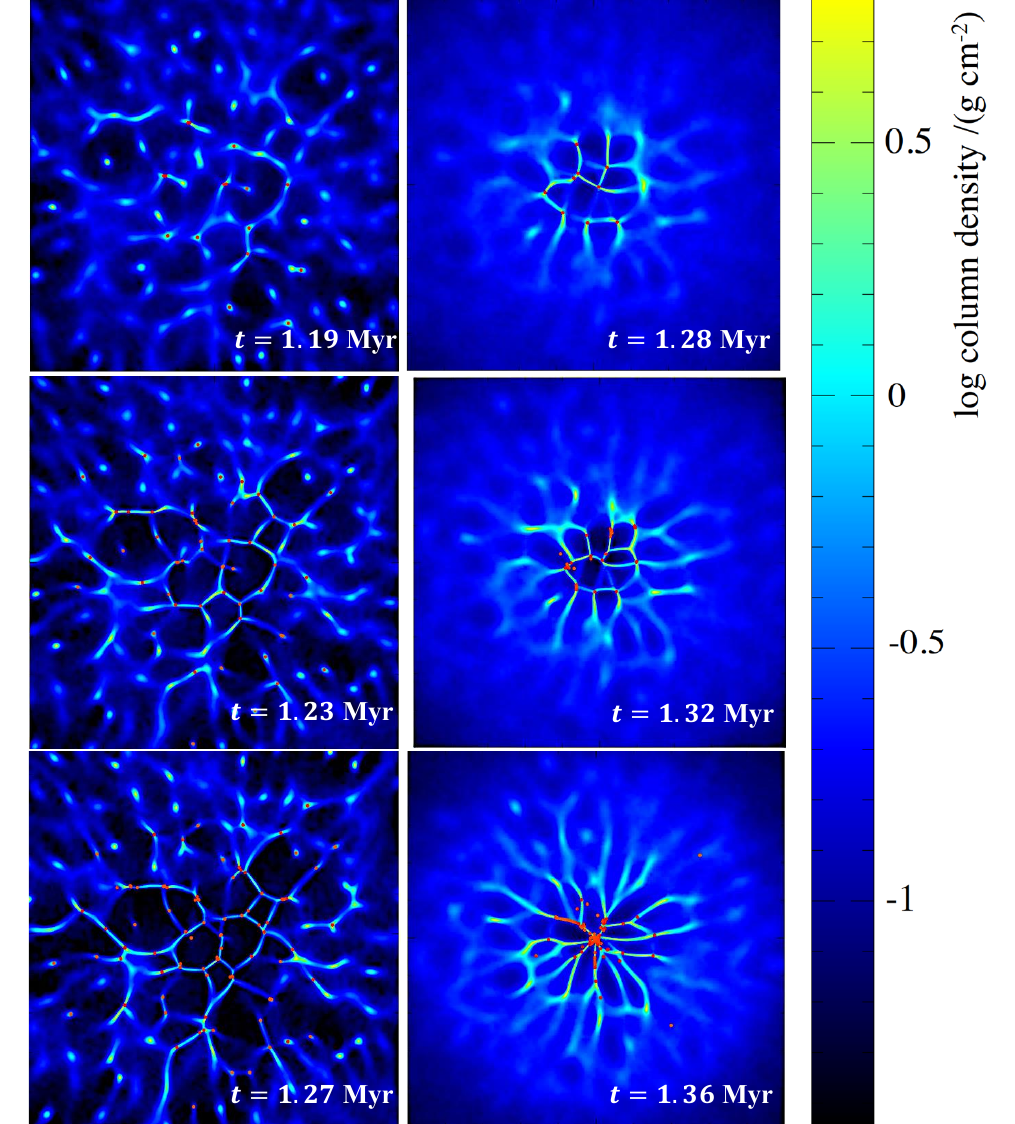}
\end{overpic}
\caption{False-colour surface-density maps looking along the $x$ axis and therefore face-on to the layer, showing the evolution of the filamentary network for collisions at $\Delta u_{\rm o}\!=\!3.6\,{\rm km\,s^{-1}}$. {\it Left panels:} no magnetic field, $B_{\rm o}\!=\!0$. {\it Right Panels:} strong magnetic field, $B_{\rm o}\!=\!5\, \upmu{\rm G}$. Each panel has dimensions $1{\rm pc}\,\times\,1{\rm pc}$. Times from the start of the collision are shown in the bottom right corner of each panel; each sequence spans $0.08\,{\rm Myr}$ and ends at $t_{10\%}$. The colour scale is logarithmic; for gas with solar composition, $\,1\,{\rm g\,cm^{-2}}$ corresponds to $\sim2.1\!\times\!10^{23}\,{\rm H_2\,cm^{-2}}$. Red dots mark the locations of star-systems. These maps have been generated using \textsc{splash} \citep{PriceD2007PASA24p159}.}
\label{Fig:Colevol}
\end{figure*}

\subsection{The effects of a sub-critical magnetic field}

At the densities ($\lesssim\rho_{\rm sink}\!=\!10^{-16}\,{\rm g\,cm^{-3}}$), and on the timescales ($\lesssim1\,{\rm Myr}$) considered here, it is reasonable to adopt ideal magneto-hydrodynamics. The gas is treated as a single fluid, which can move freely along the local field line but can only move orthogonal to the local field line if it drags the field line with it. The introduction of a magnetic field therefore has the following main effects.

Firstly, there is an additional force resisting self-gravity in directions orthogonal to the field. In effect the gravitational constant is reduced by a factor of order $(1-\beta_{\rm o}^2)$ in these directions.

Secondly, the flows that lead to the formation of filaments, then cores, and finally star-systems, tend to be aligned with the local field.

As a result, during \ACC, \LAT\ increases the strength of the axial magnetic field, $B_x$, by squeezing the field lines together, i.e. by reducing the size of the approximately circular patch where the field lines intercept the layer. Additionally the axial magnetic field increases the minimum surface-density required for fragmentation to occur, so that the \ACC\ phase takes longer.

Only during the subsequent \FRAG\ phase do the motions that collect the material of the layer into filaments create a lateral magnetic-field component, $B_{yz}$, i.e. parallel to the midplane of the layer. In the end all components of the magnetic field are increased by the convergent motions delivering mass onto and along filaments, and into cores. The magnetic field components at the highest densities tend towards equipartition (i.e. $B_{yz}\sim2^{1/2}B_x$; see Section \ref{SEC:MagField}).

\begin{figure*}
\centering
\hspace{-0.5cm}\includegraphics[width=1.1\linewidth]{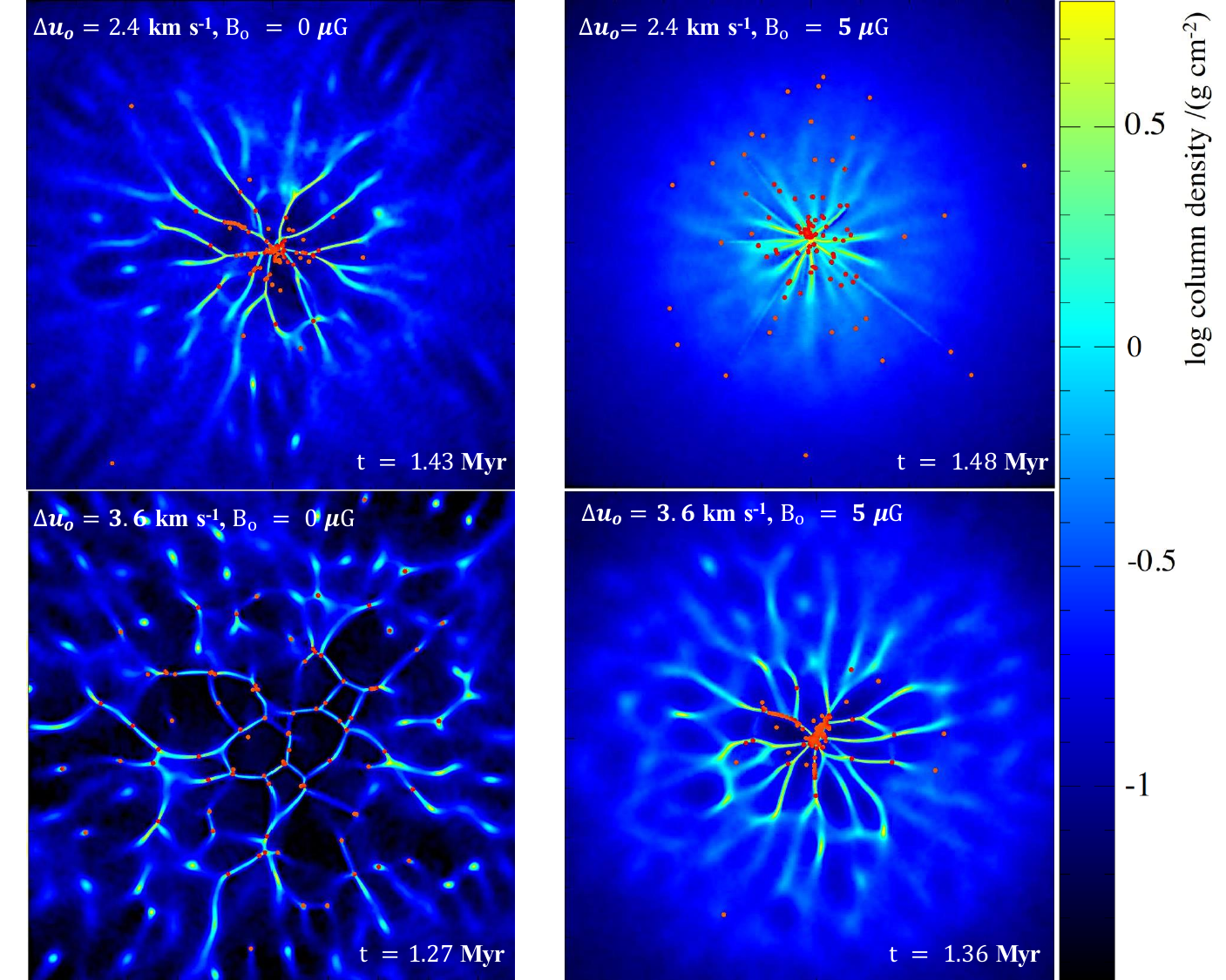}
\caption{False-colour surface-density maps at $t = t_{10\%}$, looking along the $x$ axis and therefore face-on to the layer. {\it Left panels:} no magnetic field, $B_{\rm o}\!=\!0$. {\it Right Panels:} strong magnetic field, $B_{\rm o}\!=\!5\, \upmu{\rm G}$. {\it Top Panels:} lower collision velocity ~$\Delta u_{\rm o}\!=\!2.4\,{\rm km\,s^{-1}}$. {\it Bottom Panels:} higher collision velocity $\Delta u_{\rm o}\!=\!3.6\,{\rm km\,s^{-1}}$. The initial magnetic field, $B_{\rm o}$, and the collision velocity, $\Delta u_{\rm o}$, are given in the top lefthand corner of each frame, and the value of $t_{10\%}$ in the bottom righthand corner. The colour scale is logarithmic. For gas with solar composition, $\,1\,{\rm g\,cm^{-2}}$ corresponds to $\sim2.1\!\times\!10^{23}\,{\rm H_2\,cm^{-2}}$. Red dots mark the locations of star-systems. These maps have been generated using \textsc{splash} \citep{PriceD2007PASA24p159}.}
\label{fig:ColDen}
\end{figure*}

\section{Results}\label{Sec:Results}

\subsection{The Filamentary Network}

Figure \ref{Fig:Colevol} shows a time-sequence of three surface-density maps (time increasing from top to bottom), for two different setups: the setup for the three left panels has no magnetic field; and the setup for the three right panels has a magnetic field. These maps illustrate the time evolution of the filamentary network, from the viewpoint of an observer looking along the collision axis and therefore face-on to the shock-compressed layer. The collision velocity is the same for the two setups, so the rate at which the shock compressed layer builds up,  ~$\dot{\Sigma}\sim2\rho_{\rm o}u_{\rm o}$, ~is also more-or-less the same. The differences are therefore almost entirely attributable to the fact that in one case there is no magnetic field and in the other there is.

The three maps in the lefthand column of Figure \ref{Fig:Colevol} are from an experiment with no magnetic field ($B_{\rm o}=0$) and collision velocity $\Delta u_{\rm o}\!=\!3.6\,{\rm km\,s^{-1}}$. With no magnetic field the layer fragments quickly -- i.e. before there has been significant \LAT\ -- and forms a \SPW\ of filaments. Small cores form, more-or-less independently of one another, and often at the intersections of the filaments. Small star-systems then condense out of these cores. The limited mass reservoir in an individual core is reflected in the relatively narrow mass function of the star-systems formed. There is not enough mass to form very high-mass stars, and there are few stars and therefore few violent dynamical interactions that might eject a stellar embryo, thereby cutting off accretion and creating a very low-mass star.

The three maps in the righthand column of Figure \ref{Fig:Colevol} are from an experiment with a strong magnetic field, $B_{\rm o}\!=\!5\, \upmu{\rm G}$, but the same collision velocity, $\Delta u_{\rm o}=3.6\,{\rm km\,s^{-1}}$. In this case, due to the additional magnetic pressure, the layer has to build up to a larger surface-density before it starts fragmenting, and it also fragments into filaments and cores more slowly (than in the non-magnetic case). There is therefore more time for \LAT, and since \LAT\ is non-homologous -- in the sense that the central parts of the layer contract on a shorter timescale than the outer parts -- the filaments are dragged into radial orientations. Consequently material is fed into a central monolithic core, where very high-mass star-systems can grow by competitive accretion; ~and also there are frequent dynamical interactions ejecting stellar embryos, terminating their growth and producing very low-mass star-systems. Star-systems sometimes start to condense out before their material arrives in the central core, but growth by accretion and dynamical ejection become important once they arrive in the central core.

We note that in the non-magnetic case filaments become increasingly narrow as the evolution progresses. This narrowing is predicted for an isothermal filament (strictly speaking, an isolated semi-infinite isothermal filament) by \citet{InutsukaSMiyamaS1992ApJ388p392}.

Figure \ref{fig:ColDen} shows four surface-density maps, all at $t_{10\%}$, but corresponding to different combinations of field strength $B_{\rm o}$ and collision velocity $\Delta u_{\rm o}$. The maps in the left column represent setups with no magnetic field ($B_{\rm o}=0$), and those in the right column represent setups with a strong magnetic field ($B_{\rm o}=5\,\upmu{\rm G}$). The maps on the top row are for relatively low collision velocities ($\Delta u_{\rm o}=2.4\,{\rm km\,s^{-1}}$), and those on the bottom row are for relatively high collision velocities ($\Delta u_{\rm o}=3.6\,{\rm km\,s^{-1}}$).

These maps are intended to illustrate the dichotomy between setups that evolve through a \SPW\ morphology (bottom-left panel) and produce star-systems that are initially widely distributed and relatively isolated from one another; and setups that evolve through an \HF\ morphology (the other three panels) and produce star-systems in a monolithic central star cluster, and therefore star-systems with a relatively broad mass function. In Section \ref{SEC:Locations} we discuss the consequences of this dichotomy for the spatial distribution of young star-systems.

\begin{figure*}
\includegraphics[width=1\linewidth]{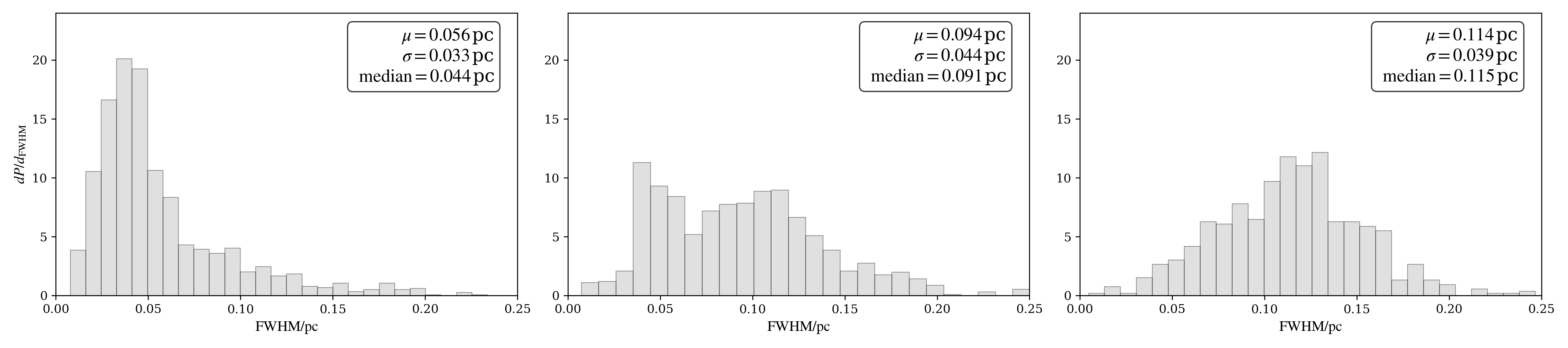}
\caption{The normalised distribution of filament widths ({\sc fwhm}) for different magnetic field strengths: ~{\it Left Panel}, no field, $B_{\rm o}\!=\!0$; ~{\it Middle Panel}, intermediate field, $B_{\rm o}\!=\!3.3\,\upmu{\rm G}$; ~{\it Right Panel}, strong field $B_{\rm o}\!=\!5\,\upmu{\rm G}$. The mean, standard deviation and median of the {\sc fwhm}, are shown in the top righthand corner of each panel. Each panel represents the results obtained from all the experiments obtained with that $B_{\rm o}$ value, i.e. three realisations for each value of $\Delta u_{\rm o}$.}
\label{fig:Width}
\end{figure*}

\subsection{Tracing Filaments}

To trace the filamentary network and extract the properties of filaments, we generate high-resolution surface-density maps like those illustrated on Figures \ref{Fig:Colevol} and \ref{fig:ColDen}. Next we convolve these maps with a Gaussian kernel having a full width at half maximum, {\sc fwhm}$\,\sim 0.01\,{\rm pc}$, corresponding to the Herschel SPIRE $250\,\upmu{\rm m}$ beam width at a distance of $\sim\!140\,{\rm pc}$ \citep[this ensures that the filament properties can be compared with observed filaments like the Taurus B211/3 filament, e.g.][]{PalmeirimPetal2013AA550A38, HowardAetal2019MN489p962}. Finally we apply the \textsc{filfinder} Python module \citep{KochERosolowskyE2015MN452p3435}: pixels exceeding the local median intensity define the filament spine and allow us to construct filament masks; radial profiles are extracted out to 0.2\,pc from the spine; each profile is then background-subtracted and fitted with a Gaussian profile to obtain the {\sc fwhm}.

\subsection{Filament {\sc fwhm}s}\label{SEC:filFWHM}

Figure \ref{fig:Width} shows the normalised distributions of filament {\sc fwhm}s for three different values of the initial magnetic field strength, $B_{\rm o}=$ $0$, $3.3$ and $5.0\,\upmu{\rm G}$. The mean {\sc fwhm} increases monotonically with increasing $B_{\rm o}$. This is (a) because the {\it lateral} field strength, $B_{yz}$, in the shock compressed layer increases with increasing $B_{\rm o}$ (see Section \ref{SEC:MagField}); and (b) because inside a filament the field tends to be aligned with the spine of the filament (see Section \ref{SEC:Alignment}), thereby giving the filament additional support orthogonal to its spine.

The distribution of filament {\sc fwhm}s for a given $B_{\rm o}$ (i.e. an individual panel on Figure \ref{fig:Width}) is generated by combining results from three independent realisations for each different collision velocity. In reality, the filament width also depends -- albeit weakly -- on the collision velocity, in the sense that the filaments tend to be slightly broader for lower collision velocities, because the magnetic field is then amplified more by \LAT\ (see Section \ref{SEC:MagField}). However, the dependence of filament width on $B_{\rm o}$ is much stronger than the dependence on $\Delta u_{\rm o}$.

Several studies \citep[e.g.][]{SmithRetal2016MN455p3640, PanopoulouGetal2017MN466p2529, PanopoulouGetal2022AA657L13, HacarAetal2018AA610A77} have explored the possibility that the appearance of a universal filament {\sc fwhm} may reflect methodological biases and/or environmental dependence. We suggest that the {\sc fwhm} of $\sim0.1$ pc frequently reported on the basis of Herschel observations of nearby clouds \citep{ArzoumanianDetal2011AA529L6, AndrePetal2016AA592A54} is unlikely to be universal, and a range of values seems more likely. In the colliding cloud setup explored here, this range reflects both the field-strength and the collision velocity. 

 A similar conclusion is drawn by \cite{PriestleyFWhitworthA2022bMN512p1407}, who study the effect of an orthogonal magnetic field on the width of an isolated radially contracting filament (i.e. a field orthogonal to the spine of the filament). They find a comparable range of filament widths, with the width increasing with increasing magnetic field  strength, and decreasing with increasing contraction speed.

\begin{figure}
\centering
\includegraphics[width = 1\linewidth]{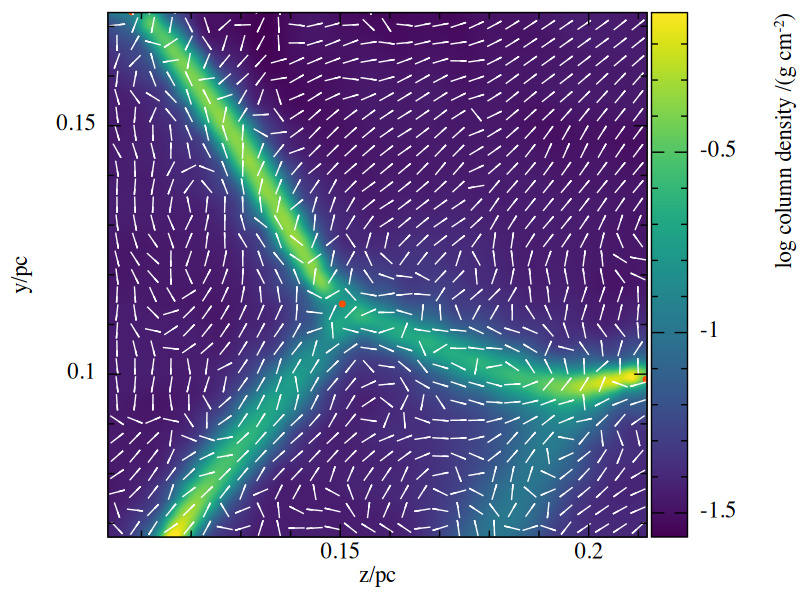}
\includegraphics[width=1\linewidth]{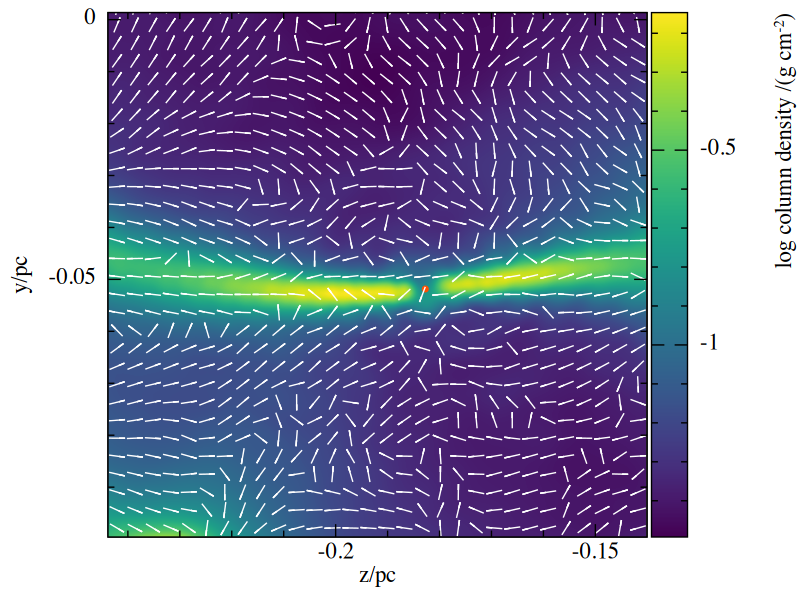}
\caption{Zoom-in false-colour surface-density maps showing the filamentary structures and magnetic field lines in the vicinity of representative star-systems: {\it Top Panel}, a star-system formed at the intersection of three filaments; {\it Bottom Panel}, a star-system formed in the middle of a filament. The colour-scale representing surface-density is logarithmic, and the white lines indicate the direction of the lateral magnetic field, $B_{yz}$. Red dots mark sink locations.}
\label{fig:Lines}
\end{figure}

\begin{figure}
\centering
\includegraphics[width=1.0\linewidth]{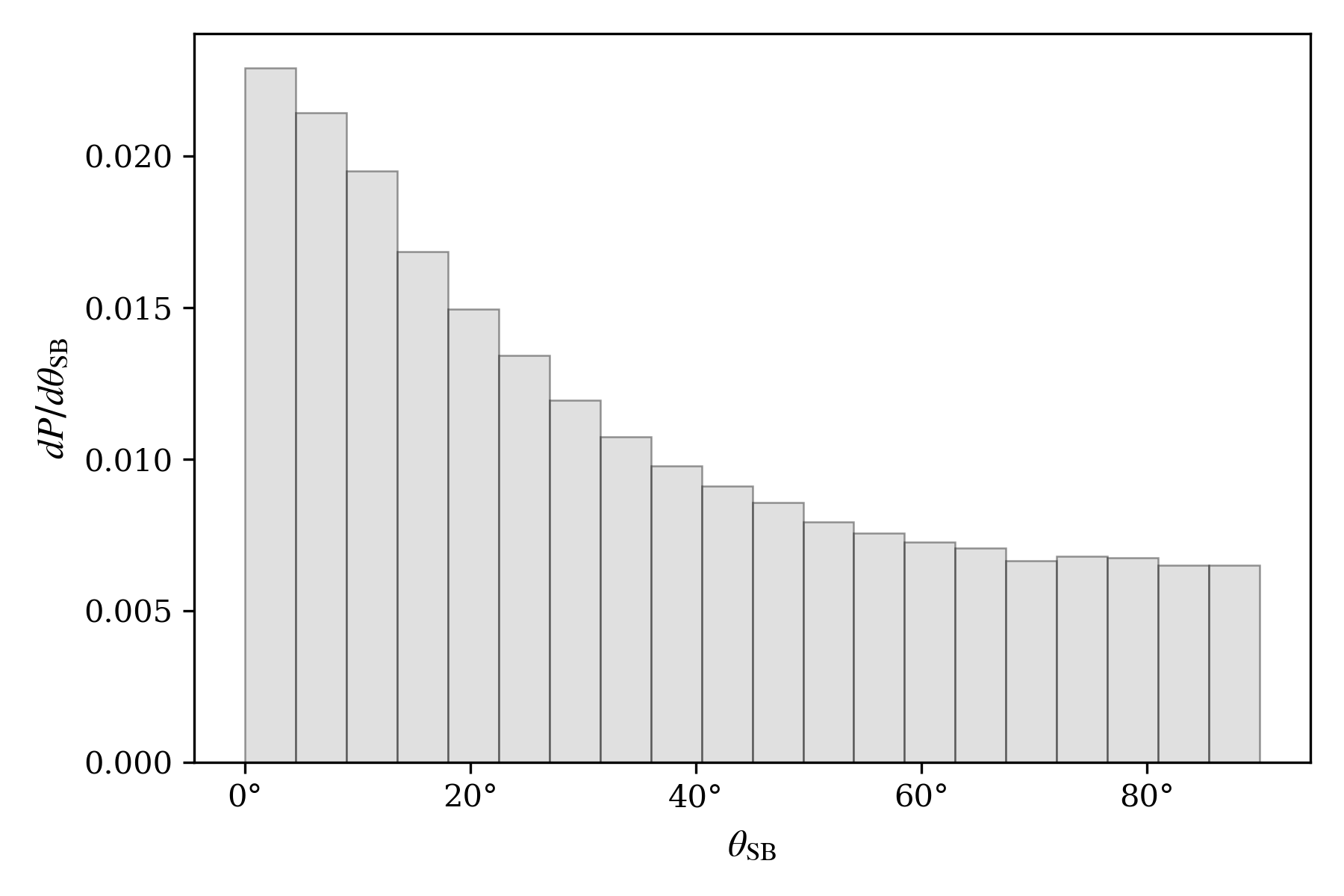}
\caption{The normalised distribution of angles, $\thSB$, between the filament spines and the local magnetic field at $t_{10\%}$. This plot derives from all the experiments (three realisations for each combination of $B_{\rm o}$ and $\Delta u_{\rm o}$).}
\label{fig:Angles}
\end{figure}

\subsection{Alignment between Filaments and the Magnetic Field}\label{SEC:Alignment}

As the shock-compressed layer contracts laterally and fragments, the gas flowing into the layer is deflected, creating a lateral magnetic field component, $B_{yz}$, which is then stretched and compressed. A stronger initial field, $B_{\rm o}$, generally results in a stronger lateral field, $B_{yz}$, in the shock-compressed layer (see Section \ref{SEC:MagField}).

Figure \ref{fig:Lines} shows the structure of the magnetic field in the vicinities of two cores, one that has formed at the intersection of three filaments, and another that has formed in the middle of a filament.  In most places the field in a filament is close to parallel with the filament spine. Under this circumstance, flow along the filament is not impeded by the Lorentz force. However, in some places, where material has accreted along a field-line and onto a filament, the field near the filament has a large component orthogonal to the spine of the filament, and the flow along the filament has had to drag the field with it, leading to a field reversal across the filament spine.

Figure \ref{fig:Angles} shows the distribution of the angles, $\thSB$, between filament spines and the local magnetic field.\footnote{In reality, projection onto the $x=0$ plane does not significantly change either the direction of the spine of a filament, or the direction of the magnetic field in a filament. Random relative orientation is therefore given by the 2D form, $d\!P\!/\!d\!\thSB\!\simeq\!1/90^{\circ},\;0^\circ\!\leq\!\thSB\!\leq\!90^{\circ}$} There is a clear preference for alignment between the spine and the field. The extra support from the aligned field then results in broader filaments, as discussed in Section \ref{SEC:filFWHM} and shown on Figure \ref{fig:Width}.

\begin{figure}
\centering
\includegraphics[width=1\linewidth]{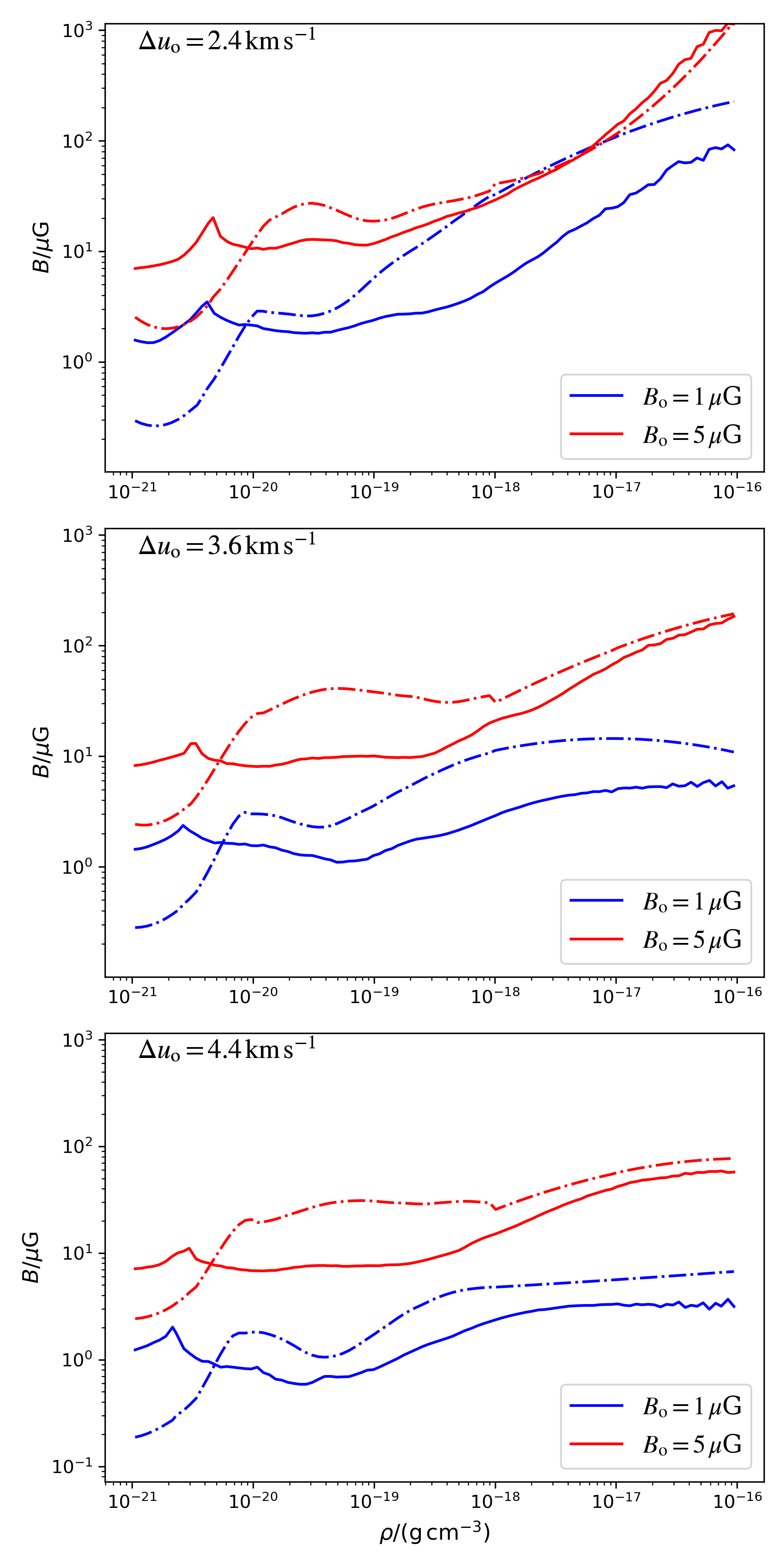}
\caption{The mean magnetic field as a function of density, at $t_{10\%}$, for collisions at different velocities: {\it Top Panel}, $\Delta u_{\rm o}\!=\!2.4\,{\rm km\,s^{-1}}$; {\it Middle Panel}, $\Delta u_{\rm o}\!=\!3.6\,{\rm km\,s^{-1}}$; {\it Bottom Panel}, $\Delta u_{\rm o}\!=\!4.4\,{\rm km\,s^{-1}}$. Results are presented separately for the axial component of the field, $B_x$ (solid lines), and for the lateral component, $B_{yz}$ (dash-dot lines). The blue curves are for $B_{\rm o}\!=\!1\,\upmu{\rm G}$, and the red curves are for $B_{\rm o}\!=\!5\,\upmu{\rm G}$. Density ranges from the initial density of the clouds ($10^{-21}\,{\rm g\,cm^{-3}}$) to the density for star-system creation ($10^{-16}\,{\rm g\,cm^{-3}}$).}
\label{fig:Bscale}
\end{figure}

\subsection{The magnetic field strength}\label{SEC:MagField}

Figure \ref{fig:Bscale} shows how the magnetic field strength depends on the density, at the end of the experiment ($t_{10\%}$), for collision velocities $\Delta u_{\rm o}=$ $2.4$, $3.6$ and $4.4\,{\rm km\,s^{-1}}$. The plotted quantity is strictly speaking an average magnetic field, $\bar{B}(\rho)$, but the standard deviation of the magnetic field at a given density is seldom large, typically $\sigma_{\log_{10}(B)}\lesssim0.2$. Results are not presented for densities $\rho\!<\!\rho_{\rm o}\!=\!10^{-21}\,{\rm g\,cm^{-3}}$, since the gas that forms the layer (and subsequently forms filaments and star-systems) never attains such low densities. Conversely, we do not show results for densities $\rho\!>\!\rho_{\rm sink}\!=\!10^{-16}\,{\rm g\,cm^{-3}}$, since there is very little of it, almost all of it has been assimilated by star-systems. There are several things to note from this plot.

Firstly, amplification of the field is quite modest, particularly for rapid collisions. Once the field has been deflected to create a lateral component and filaments have formed aligned with this lateral component, the motions collecting material into cores are predominantly along the field, and therefore do not greatly increase the field strength.

For rapid collisions, the \ACC\ and \FRAG\ phases are relatively quick, and there is little \LAT\ and hence little field compression. For example, with $B_{\rm o}=5\,\upmu{\rm G}$ and $\Delta u_{\rm o}=4.0\,{\rm km\,s^{-1}}$, the field is only amplified by a factor $\sim\!10$ over the $10^5$-fold density increase between $10^{-21}\,{\rm g\,cm^{-3}}$ and $10^{-16}\,{\rm g\,cm^{-3}}$.

For slower collisions, the \ACC\ and \FRAG\ phases are slower, and there is significant \LAT, so the field is amplified significantly by compression. For example, with $B_{\rm o}=5\,\upmu{\rm G}$ and $\Delta u_{\rm o}=2.4\,{\rm km\,s^{-1}}$, the field is amplified by a factor $\sim 200$ over the $10^5$-fold density increase between $10^{-21}\,{\rm g\,cm^{-3}}$ and $10^{-16}\,{\rm g\,cm^{-3}}$.  

Thus the rate of increase of the field with increasing density is limited to $\alpha_{_{B\rho}}\equiv d\!\ln(B)/d\!\ln(\rho)\lesssim0.5$. This exponent is significantly lower than the values for homologous 3D contraction ($\alpha_{_{B\rho}}=2/3$) or homologous 2D contraction orthogonal to the field ($\alpha_{_{B\rho}}=1$).

Secondly, the critical field strength required to inhibit star-system formation in a core is $B_{\rm sink}\simeq1.6\,{\rm mG}$ (see Equation \ref{EQN:Bsink}). If the field approaches this value, star-system formation is delayed while additional mass is accumulated by the core, and star-systems form slightly later, with slightly higher initial masses. However, from Figure \ref{fig:Bscale} we see that the field in material that might be incorporated into a star-system, i.e. material with density close to $\rho\sim10^{-16}\,{\rm g\,cm^{-3}}$, only approaches this critical field strength in the extreme case of strong initial field, $B_{\rm o}=5\,\upmu{\rm G}$, and slow collision velocity, $\Delta u_{\rm o}=2.4\,{\rm km\,s^{-1}}$ (see the red curve on the top panel of Figure \ref{fig:Bscale}). Even in this case it doesn't reach $B_{\rm sink}$.

Thirdly, the field in material that is about to be incorporated into a star-system is always approaching equipartition, $B_{yz}\sim2^{1/2}B_x$. Furthermore, where the field has not reached equipartition, the lateral component is dominant, $B_{yz}>2^{1/2}B_x$. This is the field component that is closely aligned with the filaments along which material flows into cores.

The magnetic field in gas that is about to be accreted by a star-system is higher when the collision velocity is low, because there is then a longer \ACC\ phase, and therefore more time for \LAT\ to amplify $B_x$ on its own. Consequently the subsequent lateral motions that fragment the layer and establish approximate equipartition between the field components ($B_{yz}\!\sim\!2^{1/2}B_x$) start with a larger seed field.
\begin{figure}
\centering
\includegraphics[width=1\linewidth]{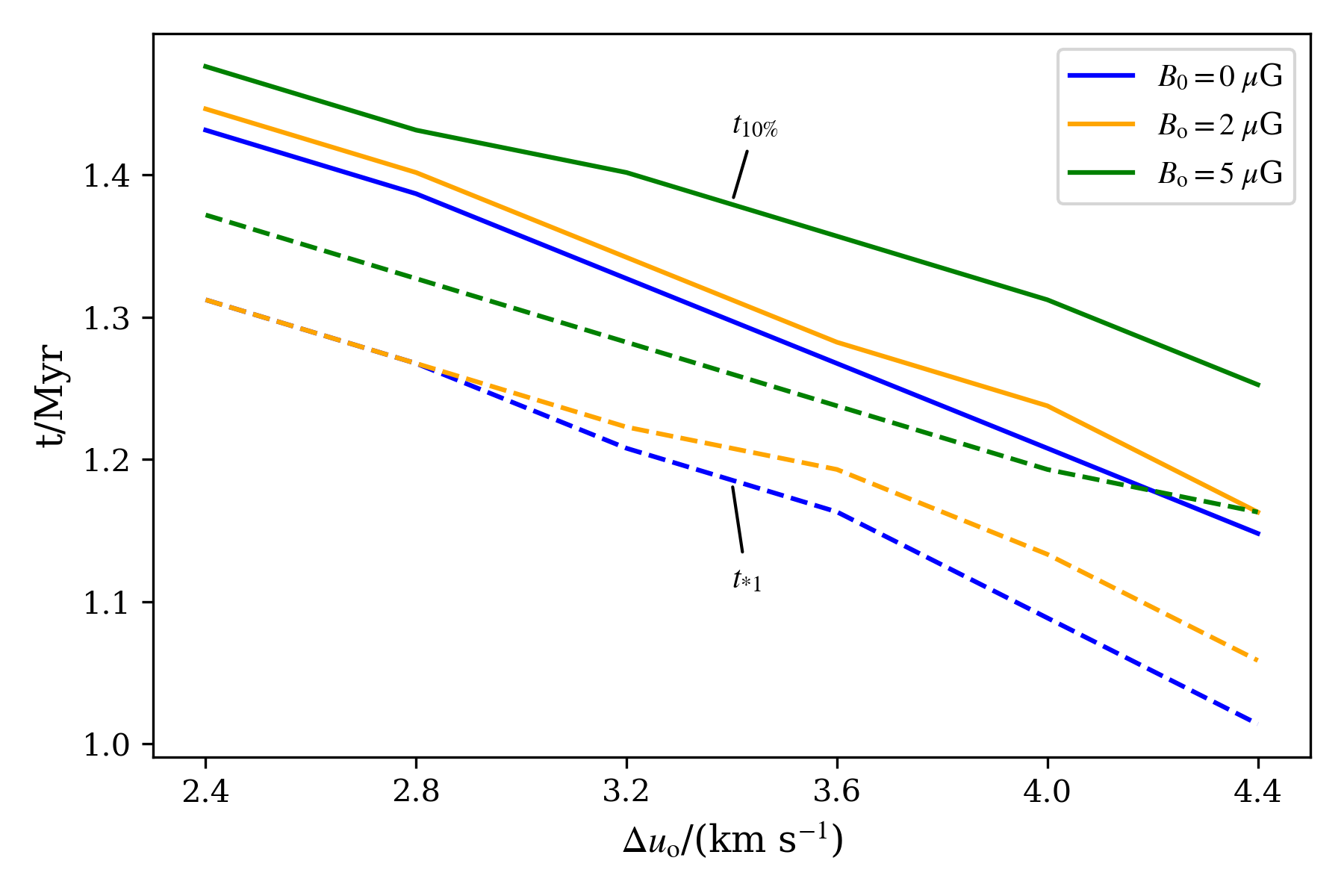}
\caption{The time at which the first star-system forms, $t_{\star1}$ ({\it dashed curves}), and the time at which $10\%$ of the mass has been converted into star-systems, $t_{10\%}$ ({\it solid curves}), as a function of collision velocity, for different initial magnetic field strengths: ~no field, $B_{\rm{o}}\!=\!0$ ({\it blue}); intermediate field, $B_{\rm{o}}\!=\!2\,\upmu{\rm G}$ ({\it orange}); ~and strong field, $B_{\rm{o}}\!=\!5\,\upmu{\rm G}$ ({\it green}). For each combination of $B_{\rm o}$ and $\Delta u_{\rm o}$ we show the average from three different realisations.}
\label{fig:firstsink}
\end{figure}

\subsection{Star-system formation timescales}

Figure \ref{fig:firstsink} shows, for different values of the initial magnetic field strength, $B_{\rm o}$, how the time at which the first star-system forms, $t_{\star1}$, and the time at which $10\%$ of the mass has been converted into star-systems, $t_{10\%}$, vary with collision velocity, $\Delta u_{\rm o}$.

$t_{\star 1}$ decreases with increasing $\Delta u_{\rm o}$ because the \ACC\ phase is shorter when the clouds collide at higher speed. ~And $t_{\star 1}$ decreases with decreasing $B_{\rm o}$ because the \FRAG\ phase is shorter when the field is weaker.

The duration of star formation, $\Delta t_{\mbox{\tiny SF}}=t_{10\%}-t_{\star 1}\simeq 0.11\pm0.01\,{\rm Myr}$, is approximately independent of both the collision velocity and the initial magnetic field strength. There are two reasons for this.

Firstly, by the time star-system creation starts at densities $\rho\gtrsim 10^{-16}\,{\rm g\,cm^{-3}}$, memory of the initial supersonic collision velocity has been erased in shocks. The dominant bulk velocities have been generated by gravitational fragmentation and are sub- or trans-sonic.

Secondly, the cores that form star-systems are all magnetically supercritical,
\begin{eqnarray}\label{EQN:Bsink}
B&<&B_{\rm sink}=2\pi G^{1/2}\rho_{\rm sink}R_{\rm sink}\sim 1.6\,{\rm mG}\,;
\end{eqnarray}
see Section \ref{SEC:MagField} and Figure \ref{fig:Bscale} for typical $B$ values at star-system creation, and \citet{ChandrasekharSFermiE1953ApJ118p116} for the expression for $B_{\rm sink}$. Once star-forming material has been collected into a core, the timescale for star formation is regulated by the interplay of self-gravity, sub- or trans-sonic turbulence, and thermal pressure. It is therefore approximately independent of $B_{\rm o}$ and $\Delta u_{\rm o}$.

\begin{figure}
\centering
\vspace{0.8cm}
\begin{overpic}[width=\linewidth]{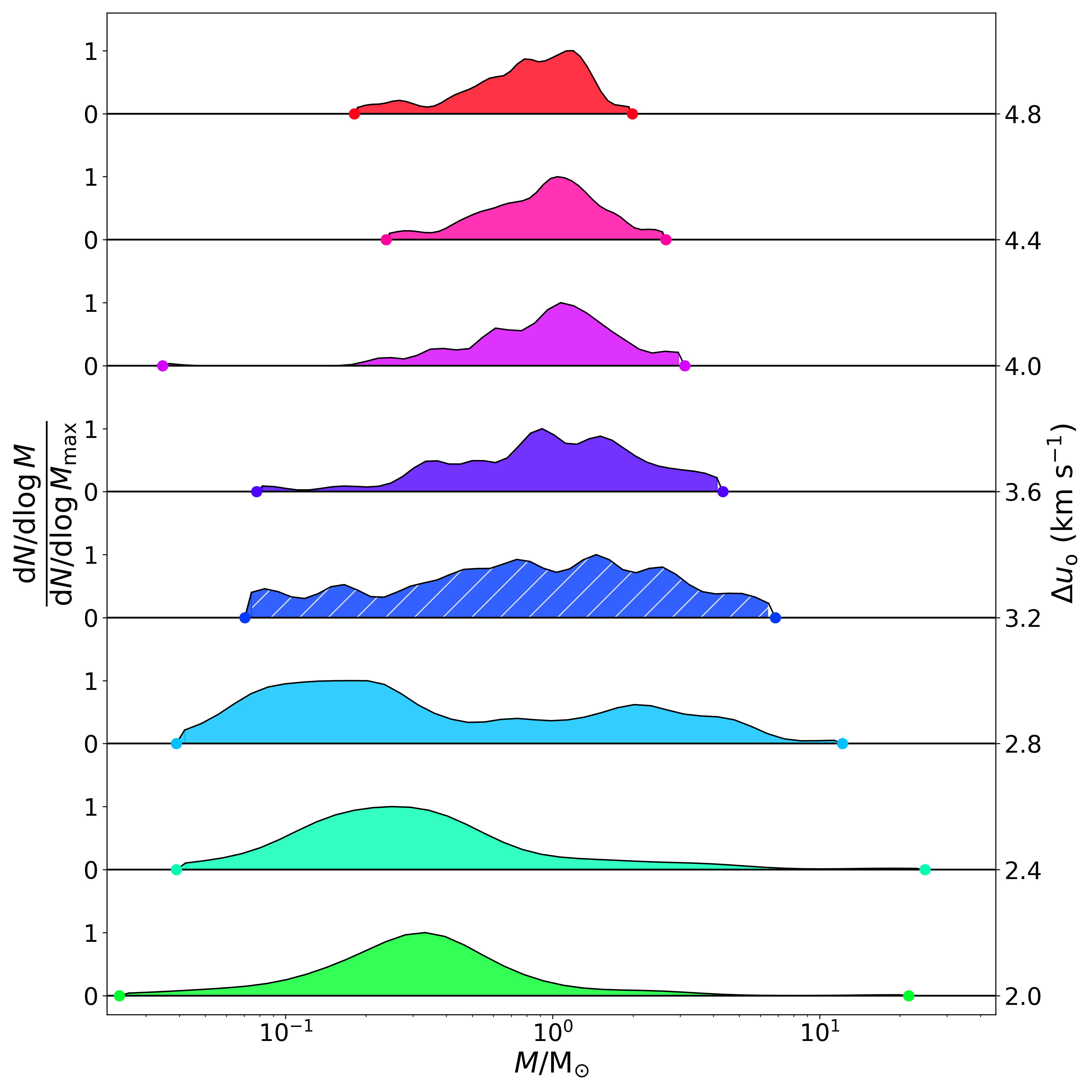}
  \put(0,102){\text{$B_{\rm o} = 0 \, \upmu$G}} 
\end{overpic}
\vspace{0.5cm}
$\,$
\begin{overpic}[width=\linewidth]{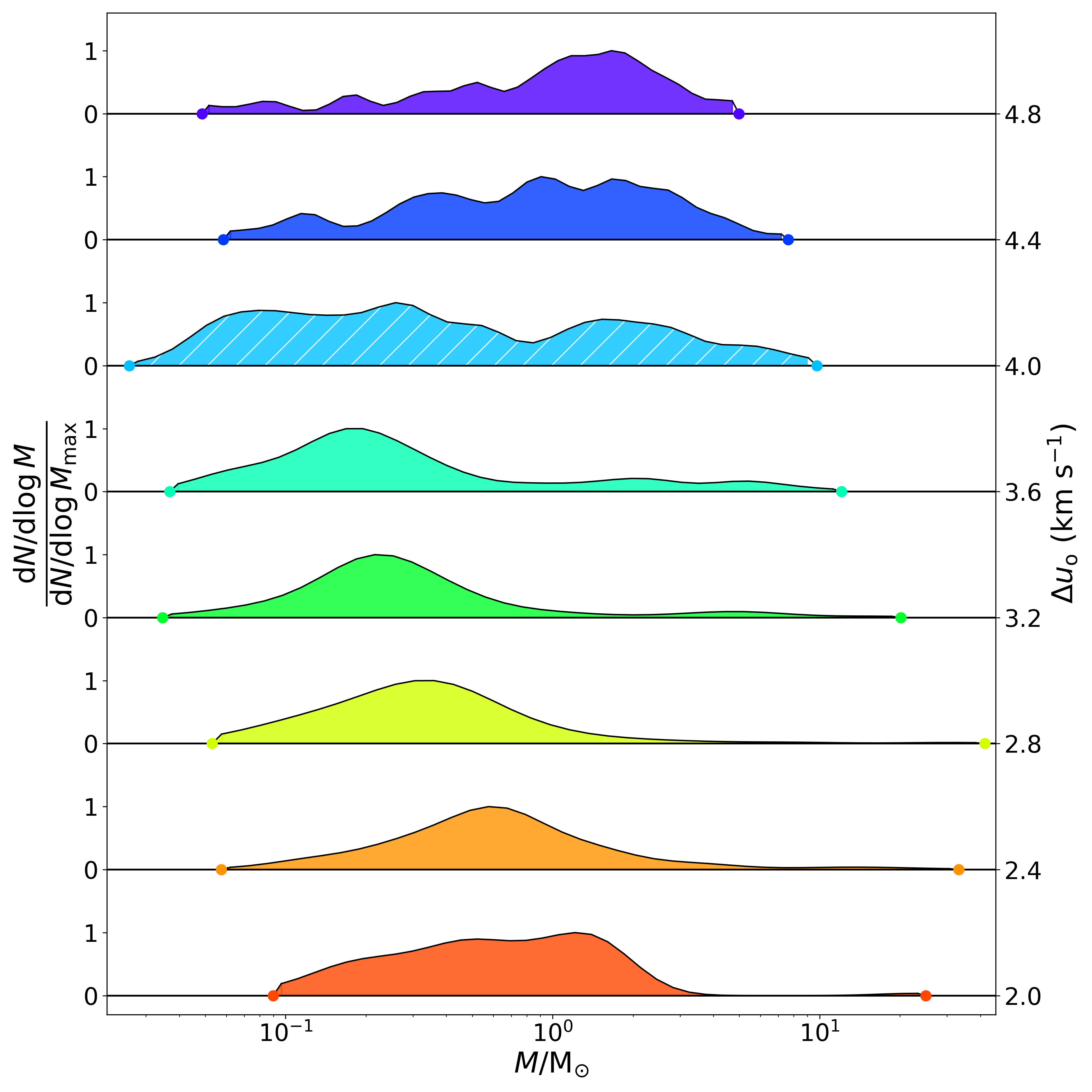}
  \put(0,102){\text{$B_{\rm o} = 5 \, \upmu$G}} 
\end{overpic}
\caption{Mass functions for six different collision velocities, $\Delta u_{\rm o}$, as indicated on the righthand axis and represented by different colours. {\it Top 8 Panels}: no magnetic field ($B_{\rm o} = 0$); {\it Bottom 8 Panels}: strong magnetic field, $B_{\rm o} = 5\, \upmu{\rm G}$. In each case there is a dot marking the minimum and maximum mass. Since each combination of $B_{\rm o}$ and $\Delta u_{\rm o}$ is only represented by $\sim 100$ star-systems, these extrema are only indicative. The distributions are normalised so that the peak is unity. The colours are selected to draw attention to the fact that the distributions with and without magnetic field are very similar, if one allows that increased magnetic field is equivalent to reduced collision velocity.}
\label{fig:Ridge}
\end{figure}

\subsection{Masses of star-systems}\label{SEC:Masses}

Figure \ref{fig:Ridge} shows the mass functions obtained with different collision velocities, $\Delta u_{\rm o}$, for cases with no magnetic field ($B_{\rm o}=0$, top 8 panels), and for cases with a relatively strong magnetic field ($B_{\rm o}=5\,\upmu{\rm G}$, bottom 8 panels). Each mass function represents the combination of three different realisations with the same $B_{\rm o}$ and $\Delta u_{\rm o}$. The value of $\Delta u_{\rm o}$ is indicated in the right margin. The distributions are coloured so as to draw attention to $(B_{\rm o},\Delta u_{\rm o})$ combinations that deliver similar mass functions, in particular the {\it purple}, {\it dark blue}, {\it pale blue}, {\it cyan} and {\it green} mass functions.

With weak (or no) magnetic field and/or relatively high collision velocity, the star formation in the shock compressed layer subscribes to a \SPW\ morphology. It delivers a relatively narrow mass function centred on $\sim 1\,{\rm M}_{_\odot}$, and few extreme masses (few very low masses due to dynamical ejection, and few very high masses due to competitive accretion), as shown at the top of Figure \ref{fig:Ridge}.

With strong magnetic field and/or relatively low collision velocity, the star formation subscribes to an \HF\ morphology. It again delivers a relatively narrow mass function, but now the mode of the mass function is shifted to lower values and depends on both $B_{\rm o}$ and $\Delta u_{\rm o}$. It shifts to lower values as $B_{\rm o}$ is decreased, or $\Delta u_{\rm o}$ is increased, as shown at the bottom of Figure \ref{fig:Ridge}. In addition there are always a few extremely massive star-systems formed by competitive accretion in the central hub, and a few very low-mass star-systems formed by dynamical ejection.

Between these limits, there is a progression that can be seen in both the non-magnetic case (top eight panels of Figure \ref{fig:Ridge}) and the strong-field case (bottom eight panels of Figure \ref{fig:Ridge}). This progression represents the transition from \HF\ star formation (at  higher $B_{\rm o}$ and/or lower $\Delta u_{\rm o}$) to \SPW\ star formation (at lower $B_{\rm o}$ and/or higher $\Delta u_{\rm o}$). 

If we focus on the non-magnetic cases (top panels of Figure \ref{fig:Ridge}), the transition from \HF\ to \SPW\ takes place around $\Delta u_{\rm o}=3.0\,{\rm km\,s^{-1}}$. Immediately above, at $\Delta u_{\rm o}=3.2\,{\rm km\,s^{-1}}$ ({\it dark blue} mass function) there is still a strong peak at $\sim 1\,{\rm M}_{_\odot}$ (similar to the {\it purple} one at $\Delta u_{\rm o}=3.6\,{\rm km\,s^{-1}}$), but there is also a growing extension to lower masses. Immediately below, at $\Delta u_{\rm o}=2.8\,{\rm km\,s^{-1}}$ ({\it pale blue} mass function) the peak at $\sim 1\,{\rm M}_{_\odot}$ has declined, and the main peak is at $\sim 0.2\,{\rm M}_{_\odot}$. At $\Delta u_{\rm o}=2.4\,{\rm km\,s^{-1}}$ ({\it cyan} mass function), the peak at $\sim 1\,{\rm M}_{_\odot}$ has almost completely disappeared, and the mass function is dominated by the peak at $\sim 0.2\,{\rm M}_{_\odot}$.

The same transition can be seen in the cases with strong magnetic field ($5\,\upmu{\rm G}$, bottom panels of Figure \ref{fig:Ridge}), but in this case the transition occurs around $\Delta u_{\rm o}=4.2\,{\rm km\,s^{-1}}$ (rather than $\Delta u_{\rm o}\sim 3.0\,{\rm km\,s^{-1}}$). This can be seen by comparing the {\it purple}, {\it dark blue}, {\it pale blue}, {\it cyan} and {\it green} mass distributions on Figure \ref{fig:Ridge}. As in the non-magnetic cases the transition marks a switch from the \HF\ morphology at low collision velocity to the \SPW\ morphology at high collision velocity.

\vspace{-0.2cm}
\begin{figure*}
\vspace{-0.2cm}
\begin{tikzpicture}
\node (A) at (0,0){\includegraphics[width=0.30\textwidth]{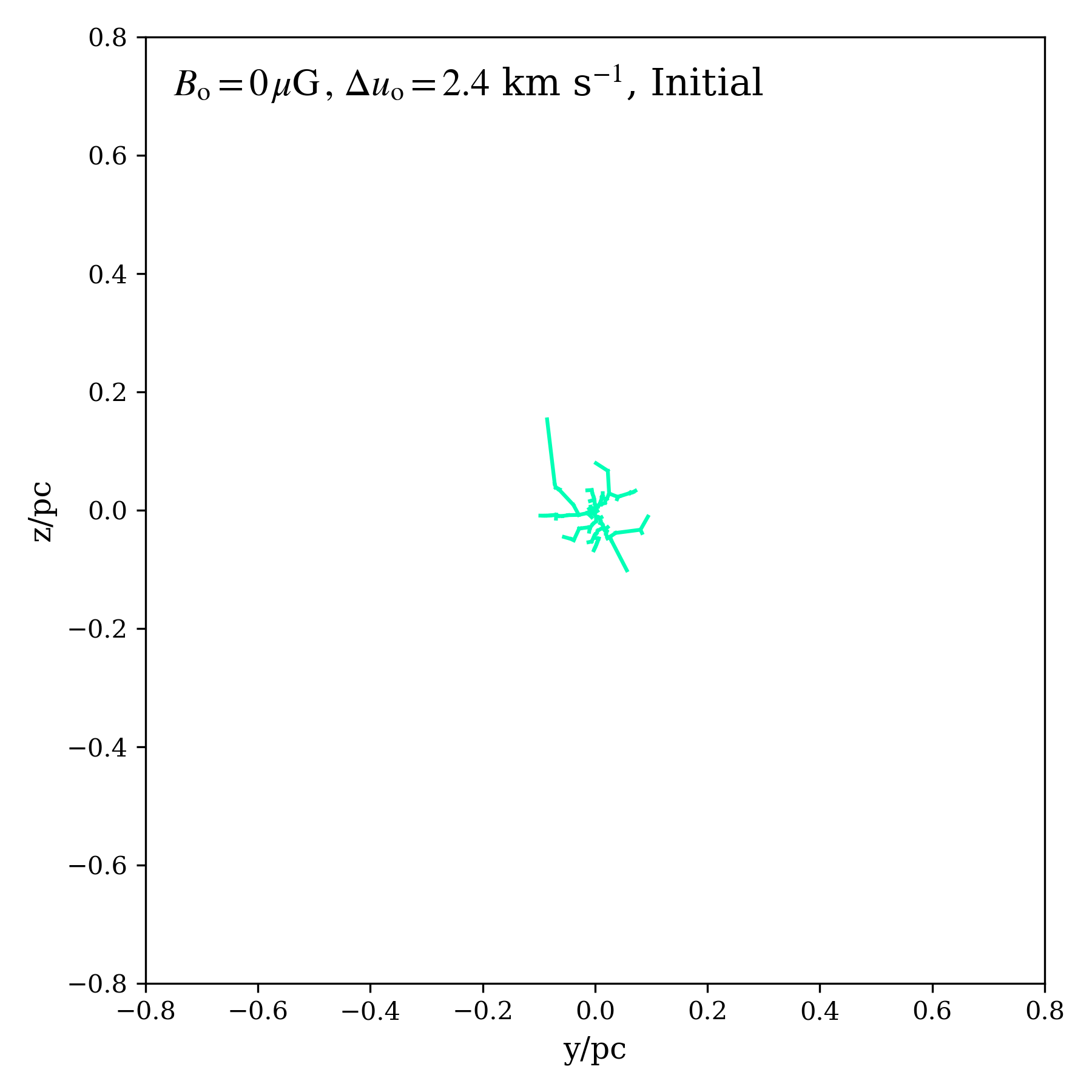}}; \node[anchor=south west] at ([yshift=-1mm]A.north west) {\small $B_{\rm o} = 0\,\upmu$G};
\end{tikzpicture}
\includegraphics[width=0.30 \linewidth]{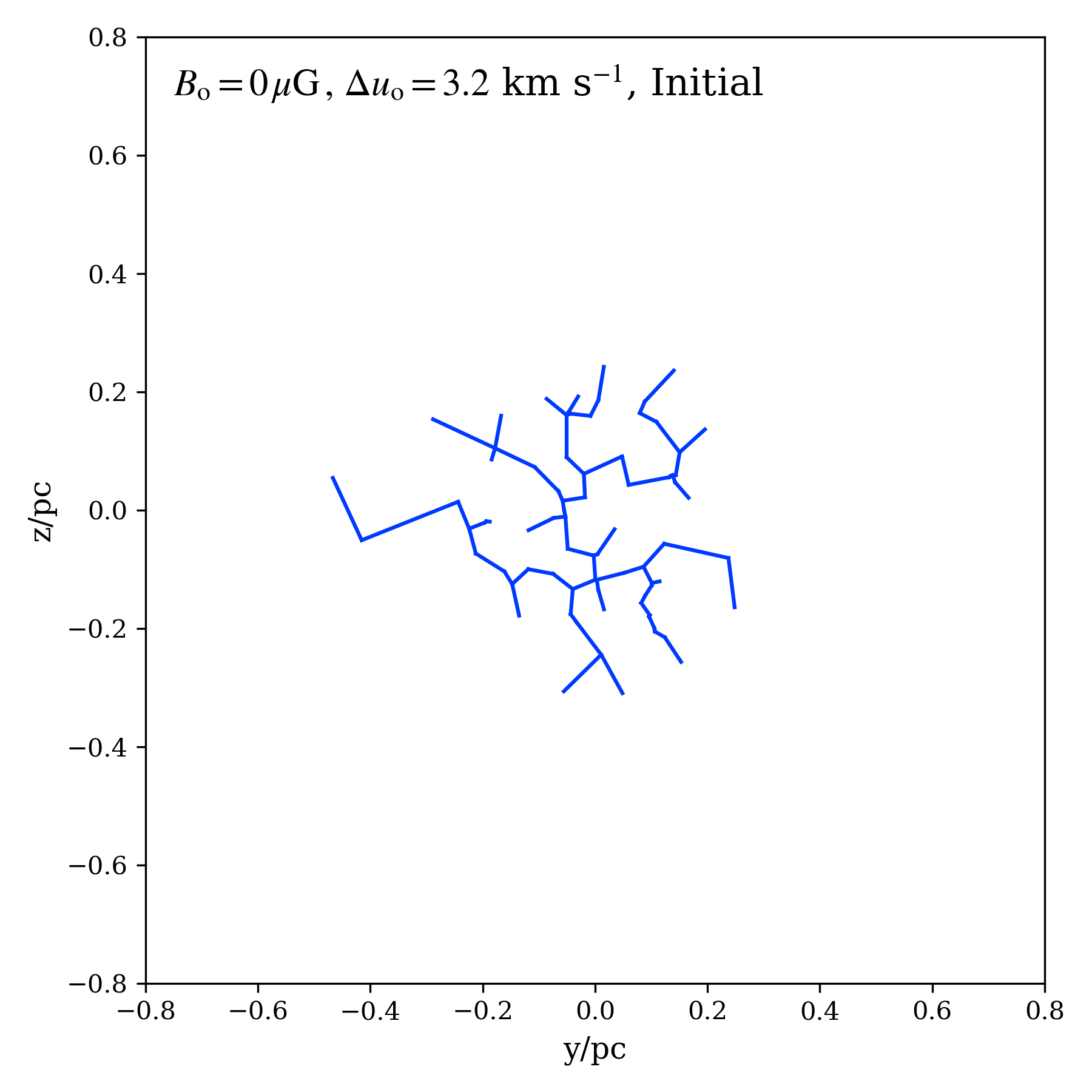}
\includegraphics[width=0.30 \linewidth]{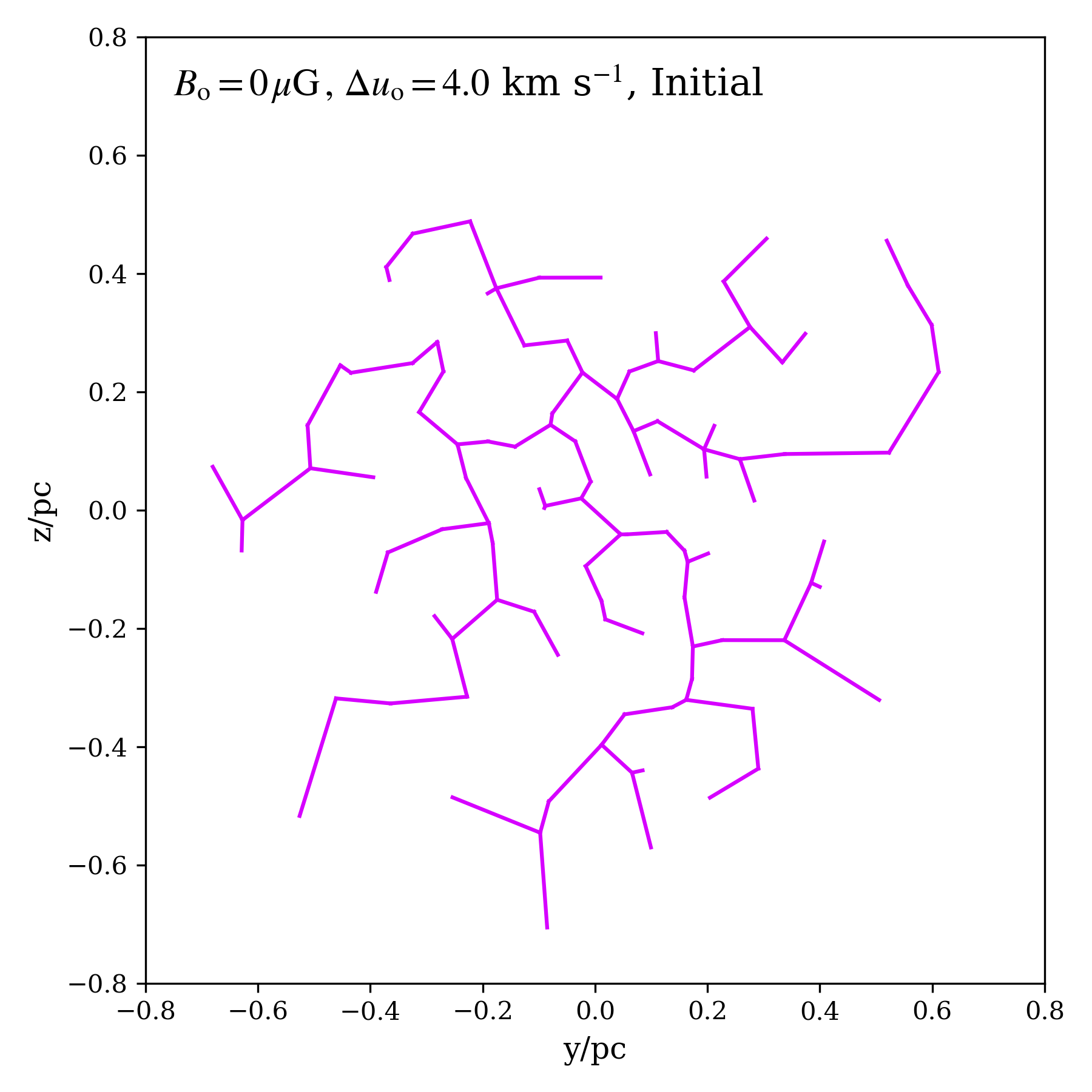}    
\includegraphics[width=0.30 \linewidth]{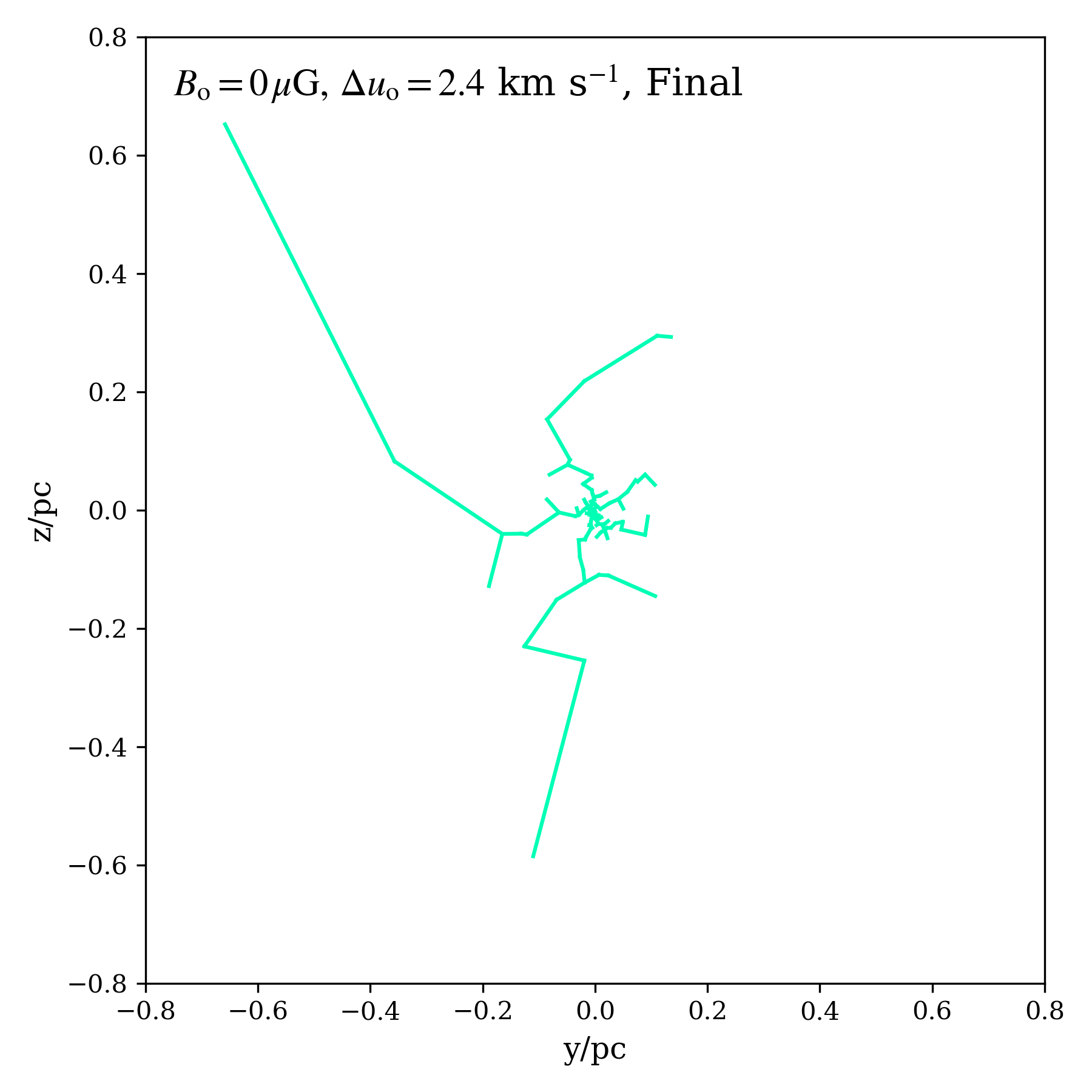}
\includegraphics[width=0.30 \linewidth]{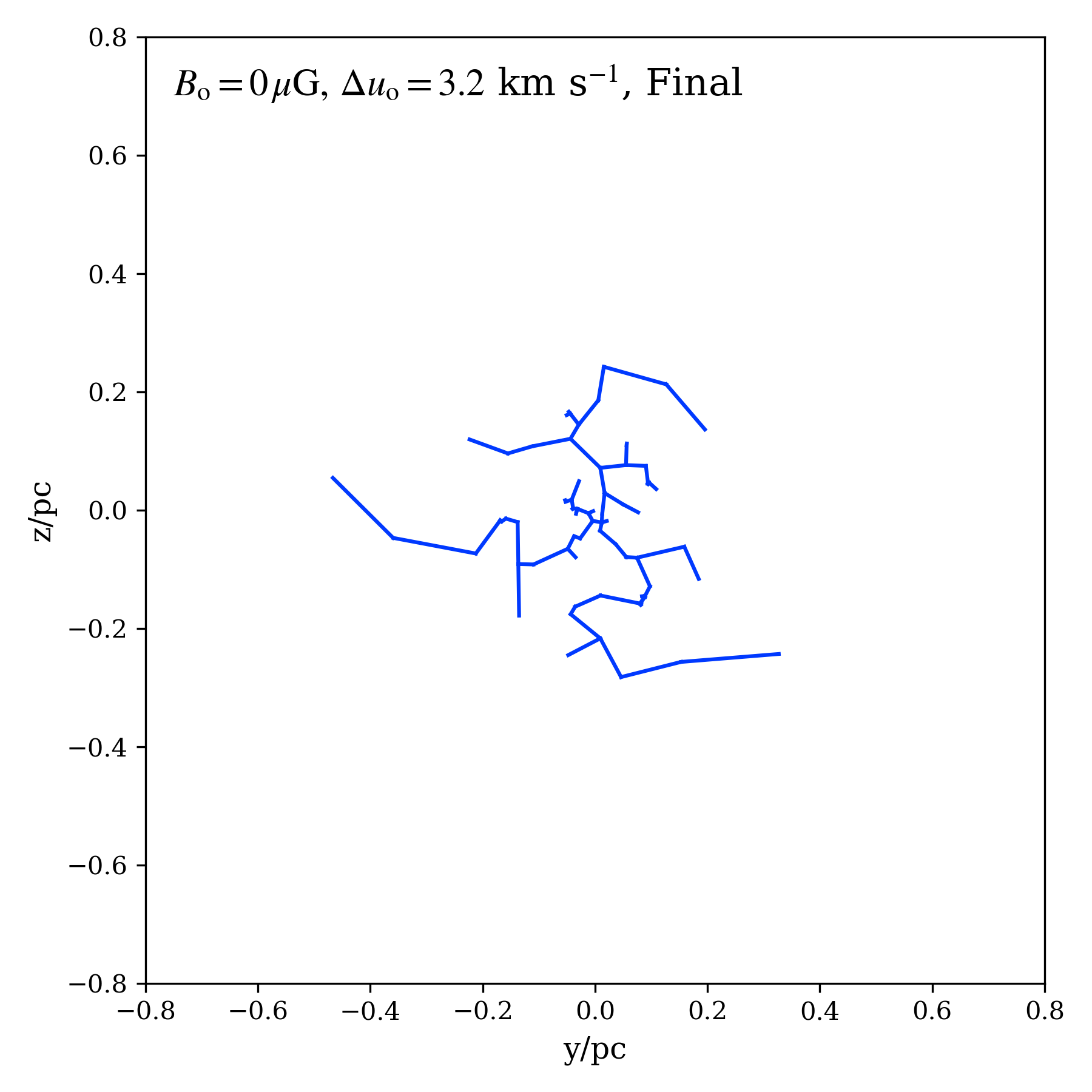}
\includegraphics[width=0.30 \linewidth]{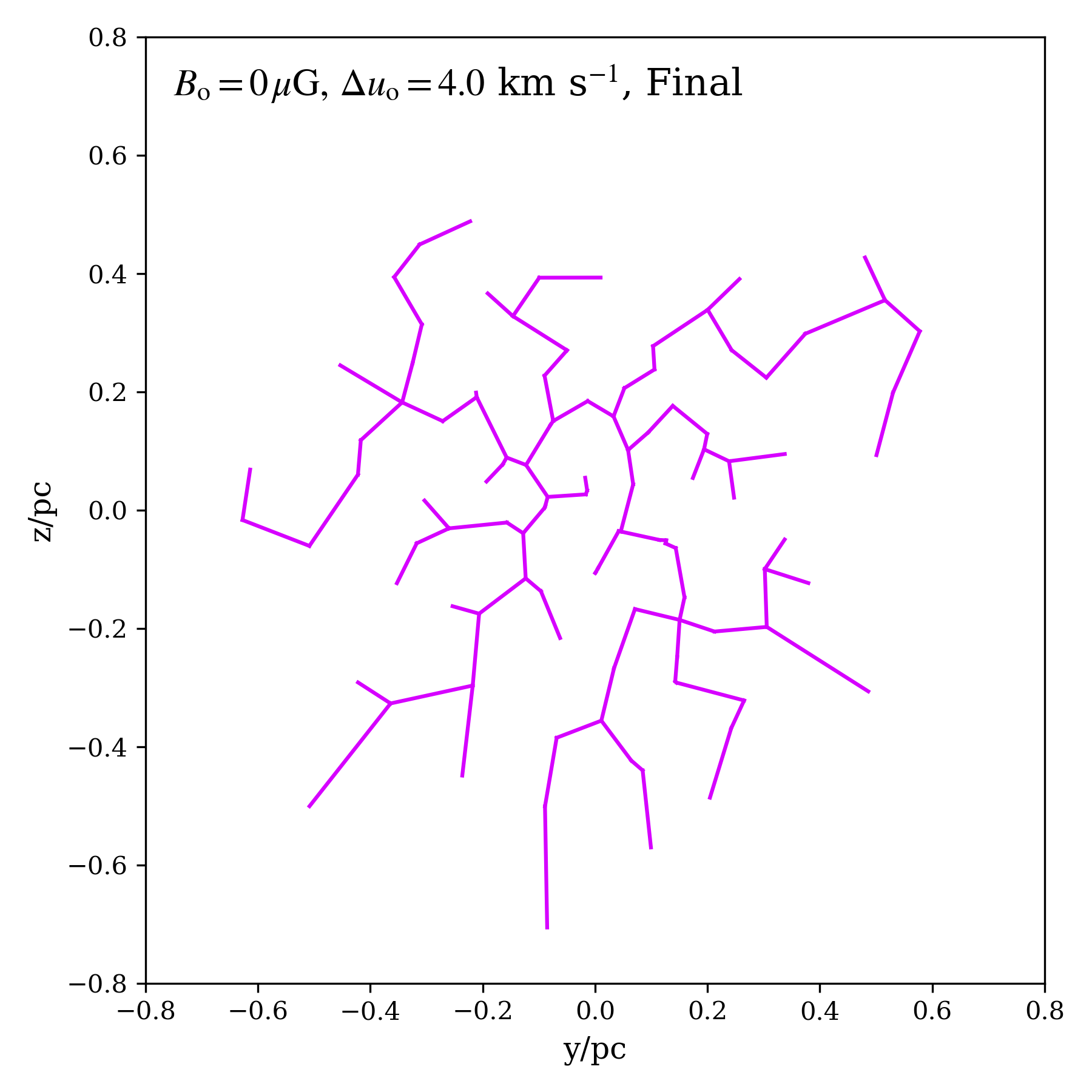}
\begin{tikzpicture}
\node (B) at (0,0){\includegraphics[width=0.30\textwidth]{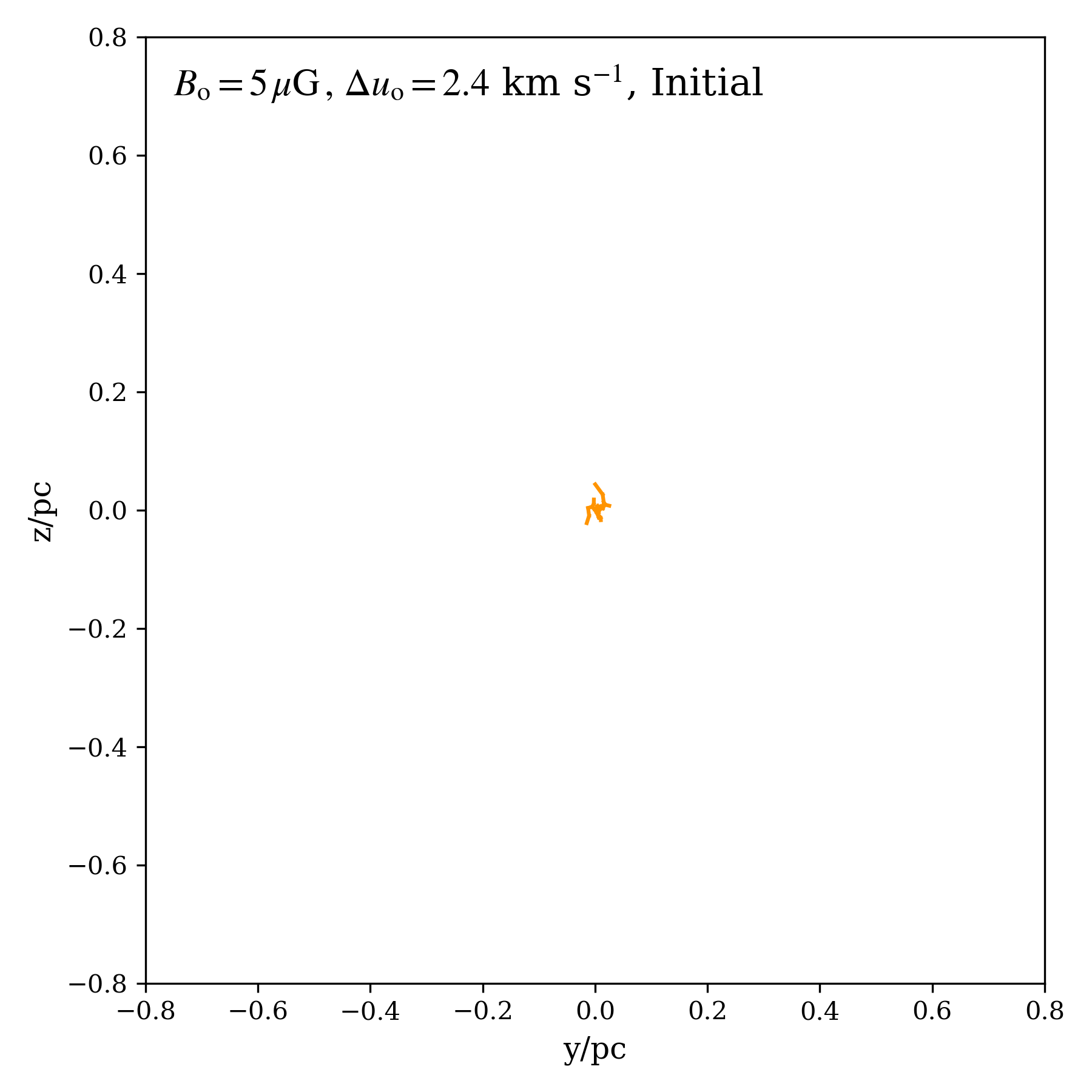}}; \node[anchor=south west] at ([yshift=-1mm]B.north west) {\small $B_{\rm o} = 5\,\upmu$G};
\end{tikzpicture}
\includegraphics[width=0.30 \linewidth]{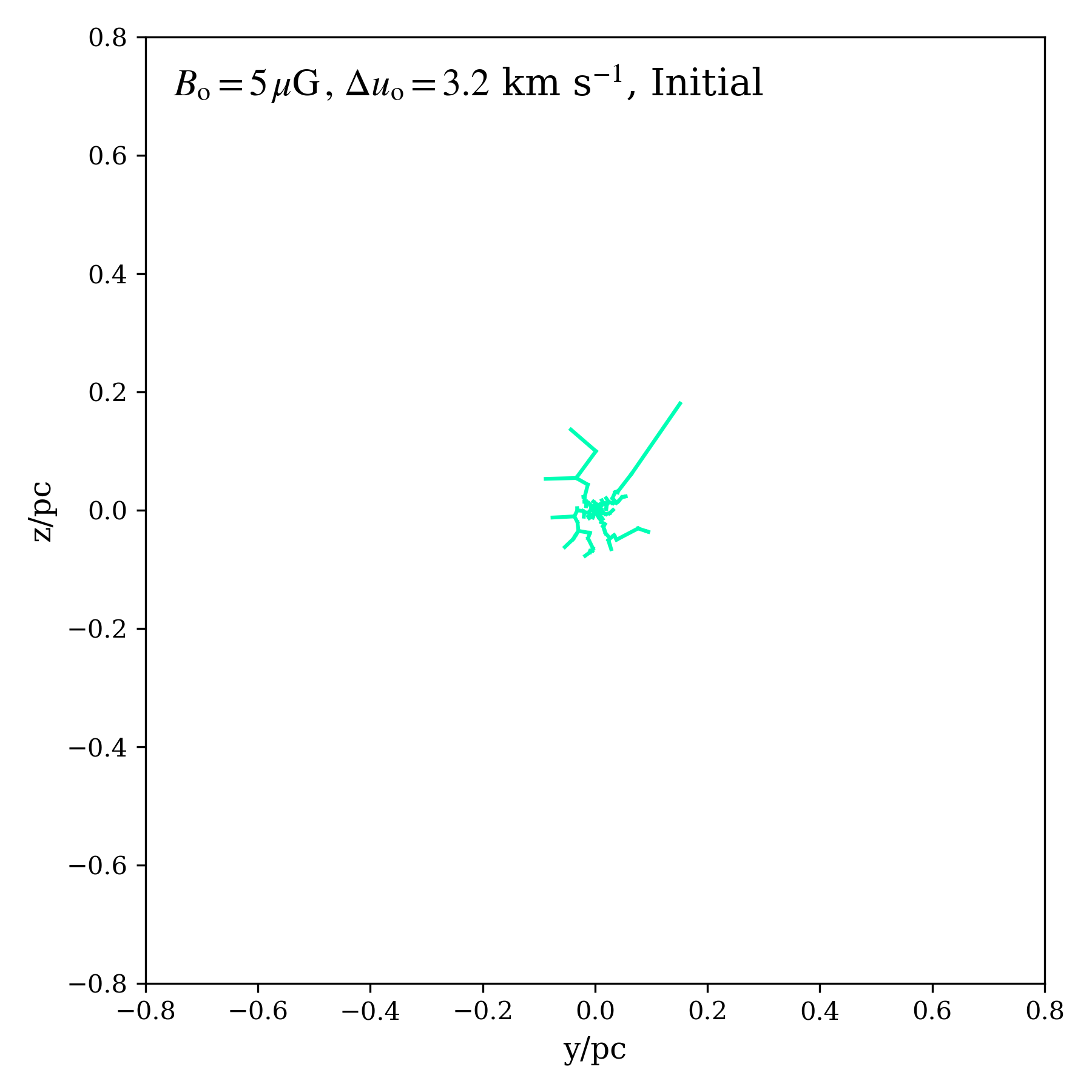}
\includegraphics[width=0.30 \linewidth]{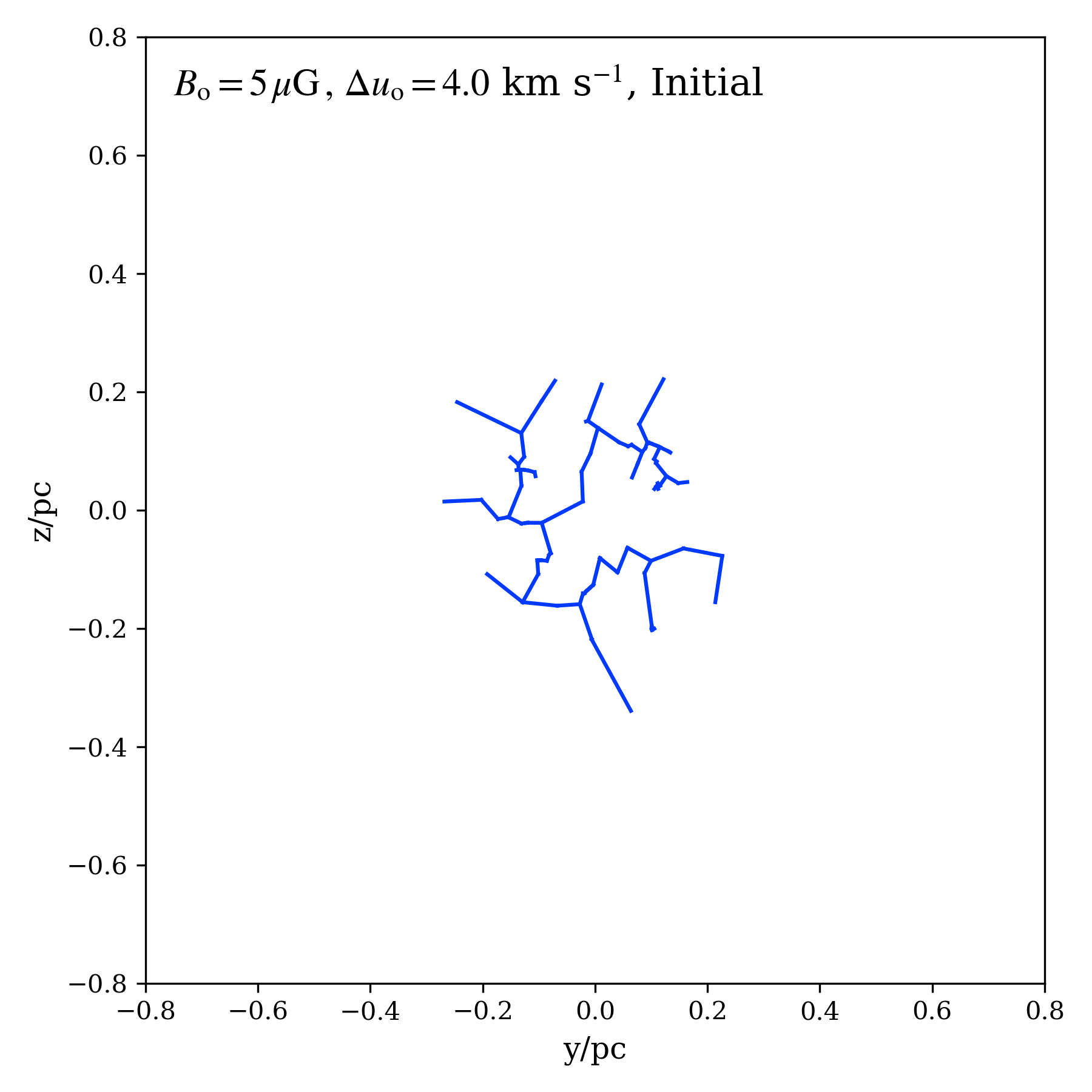}
\includegraphics[width=0.30 \linewidth]{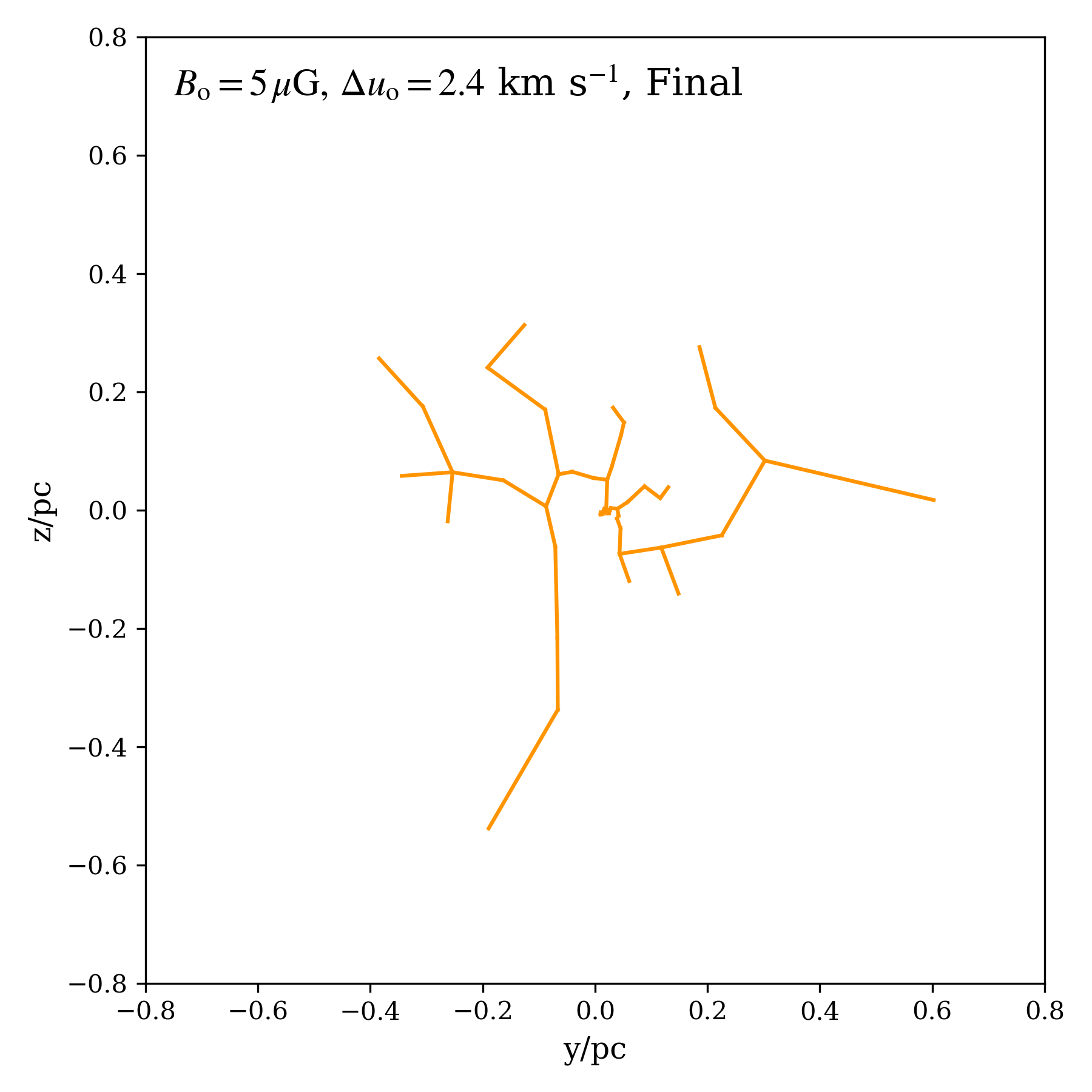}
\includegraphics[width=0.30 \linewidth]{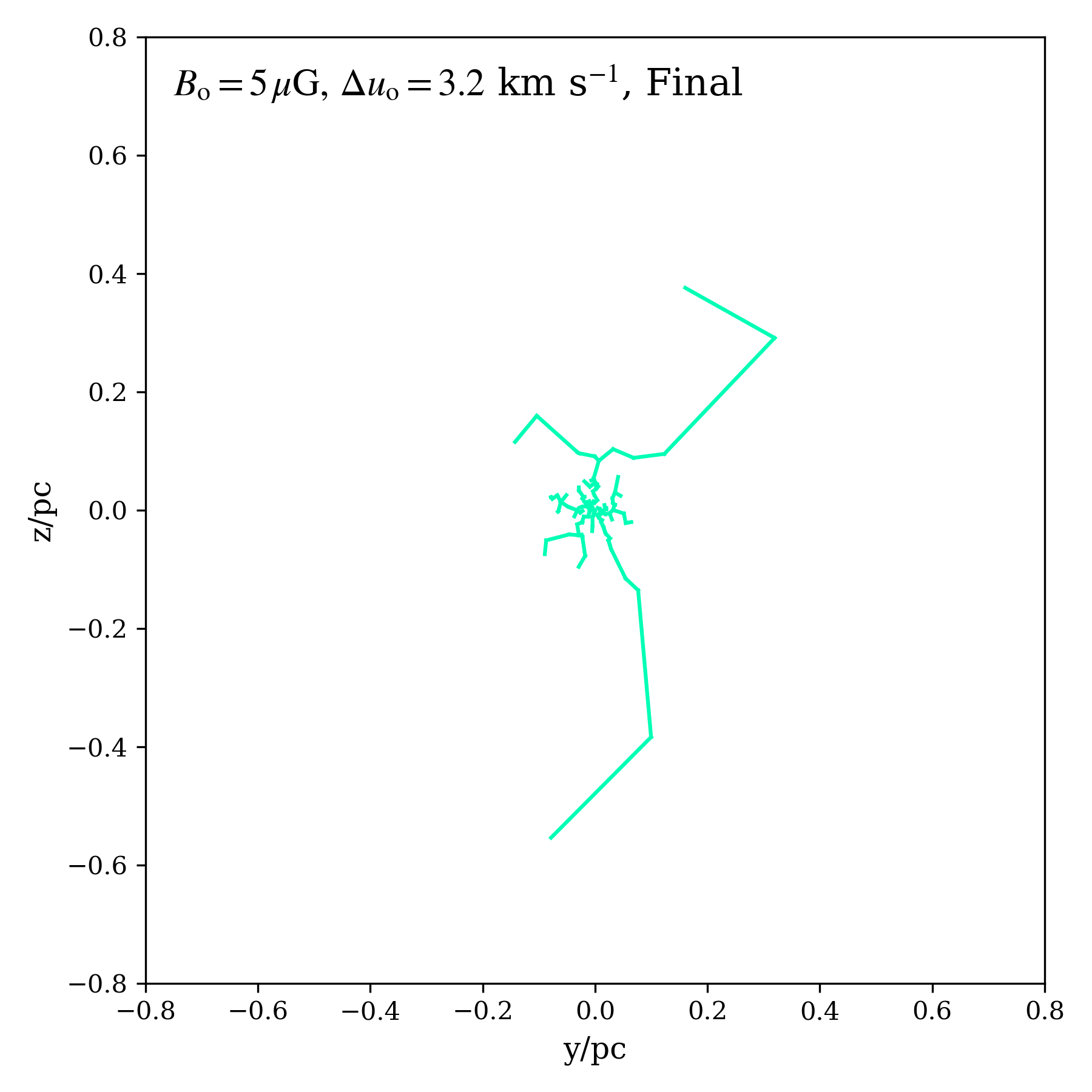}
\includegraphics[width=0.30 \linewidth]{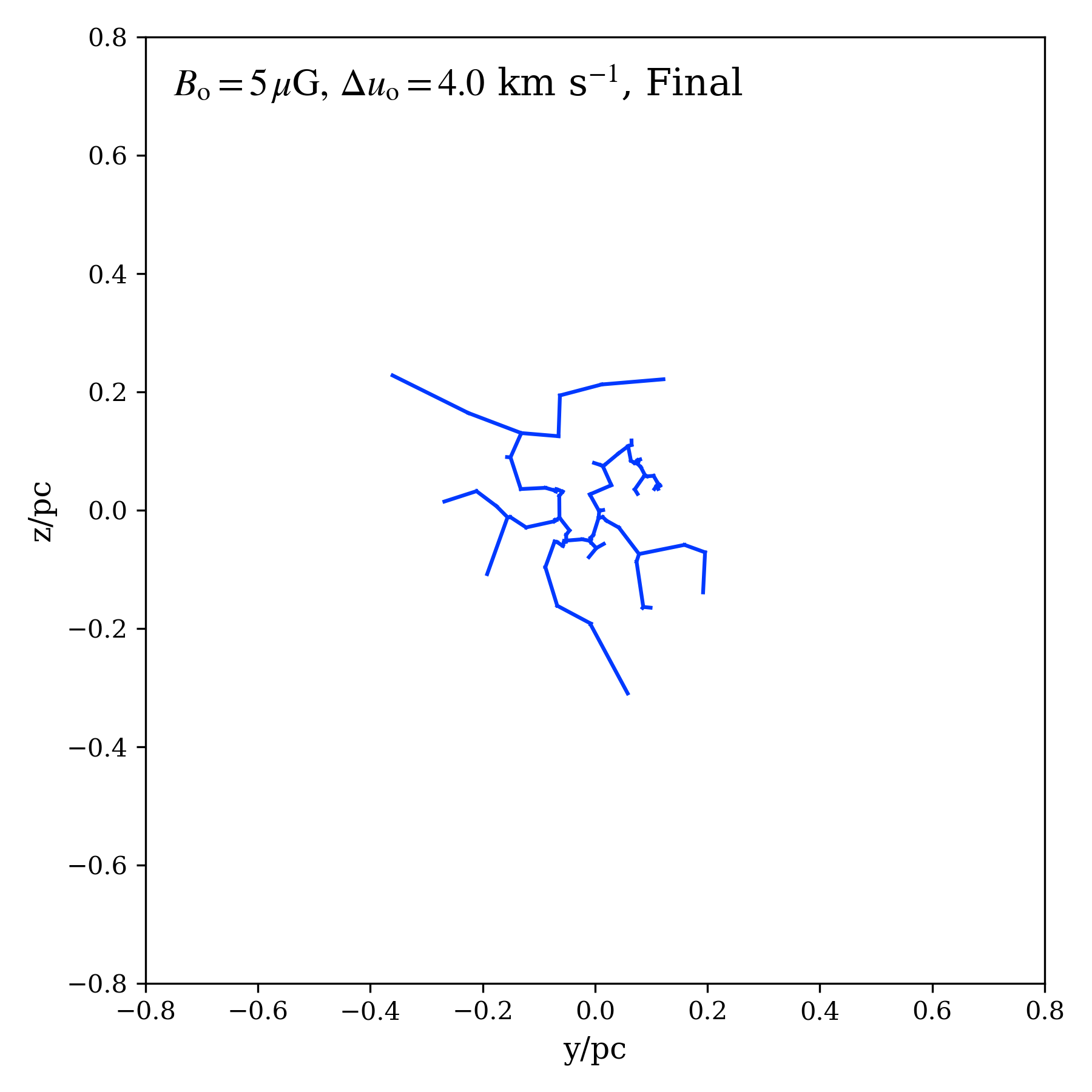}
\caption{Minimum spanning trees (MSTs) for star-system positions projected onto the $x\!=\!0$ plane. ~{\it First \& Second Rows}: initial and final star-system positions with no magnetic field ($B_{\rm o}=0$). {\it Third \& Fourth Rows}: initial and final star-system positions with strong magnetic field, $B_{\rm o}=5\,\upmu{\rm G}$. ~ {\it Left Column}: $\Delta u_{\rm o} = 2.4\,{\rm km\,s^{-1}}$; {\it Middle Column}: $\Delta u_{\rm o} = 3.6\,{\rm km\,s^{-1}}$; {\it Right Column}: $\Delta u_{\rm o} = 4.0\,{\rm km\,s^{-1}}$. The MSTs are coloured so as to draw attention to $(B_{\rm o},\Delta u_{\rm o})$ combinations that have similar initial star-system MSTs {\it and} similar final star-system MSTs.}
\label{fig:MSTall}
\end{figure*}

\subsection{The spatial distribution of star-systems}\label{SEC:Locations}

Figure \ref{fig:MSTall} shows a selection of Minimum Spanning Trees (MSTs) for the positions of the star-systems projected onto the $x=0$ plane. The top six panels are for non-magnetic cases ($B_{\rm o}=0$), and the bottom six panels are for strong magnetic field cases ($B_{\rm o}=5\,\upmu{\rm G}$). Panels in the left column represent relatively slow collisions, with $\Delta u_{\rm o}=2.4\,{\rm km\,s^{-1}}$; panels in the central column represent intermediate-velocity collisions, with $\Delta u_{\rm o}=3.6\,{\rm km\,s^{-1}}$; and panels in the right column represent relatively fast collisions, with $\Delta u_{\rm o}=4.0\,{\rm km\,s^{-1}}$. For each $(B_{\rm o},\Delta u_{\rm o})$ combination there is an upper panel showing the MST of the initial (i.e. formation) positions, and a lower panel showing the MST of the final ($t_{10\%}$) positions. The initial positions are {\it de facto} not simultaneous, whereas the final positions are simultaneous. 


In an {\sc extreme} \HF\ setup, the collision is so slow, and/or the field so strong, that the \ACC\ and \FRAG\ phases only get started after there has been considerable \LAT. Because \LAT\ is non-homologous (i.e. the centre of the shock-compressed layer contracts laterally on a shorter timescale than the outer parts) the filaments are dragged into radial orientations, and the shock compressed gas is fed into the central hub, before many star-systems condense out. Once they do, competitive accretion creates a few very massive star-systems. At the same time there are violent dynamical ejections creating a diaspora of very low-mass star-systems. The $[B_{\rm o},\,\Delta u_{\rm o}]=$ $[5\,\upmu{\rm G},\,2.4\,{\rm km\,s^{-1}}]$ case, with {\it orange} MSTs on Figure \ref{fig:MSTall}, is an example of an {\sc extreme} \HF\ setup. The spatial distribution of star-systems at formation is extremely compact, and the MST at $t_{10\%}$ is very extended due to dynamical ejections.

In a {\sc moderate} \HF\ setup, the collision is sufficiently slow, and/or the field sufficiently strong, that the \ACC\ and \FRAG\ phases proceed quite slowly, and there is sufficient time for some \LAT\ before they are complete.  The filaments are dragged into radial orientations, so that gas and forming star-systems are funnelled into a central hub. Here they continue to grow by accretion, unless they are ejected dynamically. The $[B_{\rm o},\,\Delta u_{\rm o}]=$ $[0\,\upmu{\rm G},\,2.4\,{\rm km\,s^{-1}}]$ and $[5\,\upmu{\rm G},\,3.2\,{\rm km\,s^{-1}}]$ cases, with {\it green} MSTs on Figure \ref{fig:MSTall}, are examples of {\sc moderate} \HF\ setups. The MST at $t_{10\%}$ is more extended than the MST at formation, by virtue of the dispersed positions of the dynamically ejected star-systems, but the difference is not as marked as for the ({\it orange}) {\sc extreme} \HF\ setup.

In a {\sc moderate} \SPW\ setup, the collision is sufficiently fast, and/or the field sufficiently weak, that the \ACC\ and \FRAG\ phases proceed quite rapidly. There is enough time for a little \LAT, but not enough for the filaments to be dragged into radial orientations. Consequently there is a \SPW\ of intersecting filaments, which spawn an array of star-systems. The lateral extent of the array is reduced somewhat by \LAT. The $[B_{\rm o},\,\Delta u_{\rm o}]=$ $[0\,\upmu{\rm G},\,3.2\,{\rm km\,s^{-1}}]$ and $[5\,\upmu{\rm G},\,4.0\,{\rm km\,s^{-1}}]$ cases, with {\it blue} MSTs on Figure \ref{fig:MSTall}, are examples of {\sc moderate} \SPW\ setups. The spatial distribution of star-systems at formation and at $t_{10\%}$ are similar, because there are not many dynamical ejections.

In an {\sc extreme} \SPW\ setup, the collision is so fast, and the field is so weak, that the \ACC\ and \FRAG\ phases are completed before there is enough time for significant \LAT, and therefore before there is time to decrease the lateral extent of the shock-compressed layer, or to drag the filaments into radial orientations. Consequently \FRAG\ produces an extended \SPW\ of intersecting filaments, which in turn spawn a widely dispersed array of star-systems. The $[B_{\rm o},\,\Delta u_{\rm o}]=$ $[5\,\upmu{\rm G},\,4.0\,{\rm km\,s^{-1}}]$ case, with {\it purple} MSTs on Figure \ref{fig:MSTall}, is an example of an {\sc extreme} \SPW\ setup. The spatial distributions of star-systems at formation and at $t_{10\%}$ are similar. 

Thus there is always a progression from ~(a) an extreme \HF\ morphology producing a monolithic cluster with the formation of star-systems confined to a very compact initial MST, but then high ejection speeds producing an extended final MST at $t_{10\%}$ ({\it orange plots}); through ~(b) (ii) a less extreme \HF\ morphology with the formation of star-systems in a moderately compact initial MST and a few ejections extending the final MST with a few straggly arms at $t_{10\%}$ ({\it green plots}); on to ~(c) a relatively compact \SPW\ morphology with a moderately extended initial MST that has not expanded much by $t_{10\%}$ (the final MST), because there have not been many dynamical ejections ({\it blue plots}); and  finally on to ~(d) a broad \SPW\ morphology producing an array of isolated star-systems, where again the MST does not change much between formation (initial MST) and $t_{10\%}$ (final MST), because there are few energetic dynamical ejections ({\it purple plots}).

\begin{figure}
\centering
\includegraphics[width=1\linewidth]{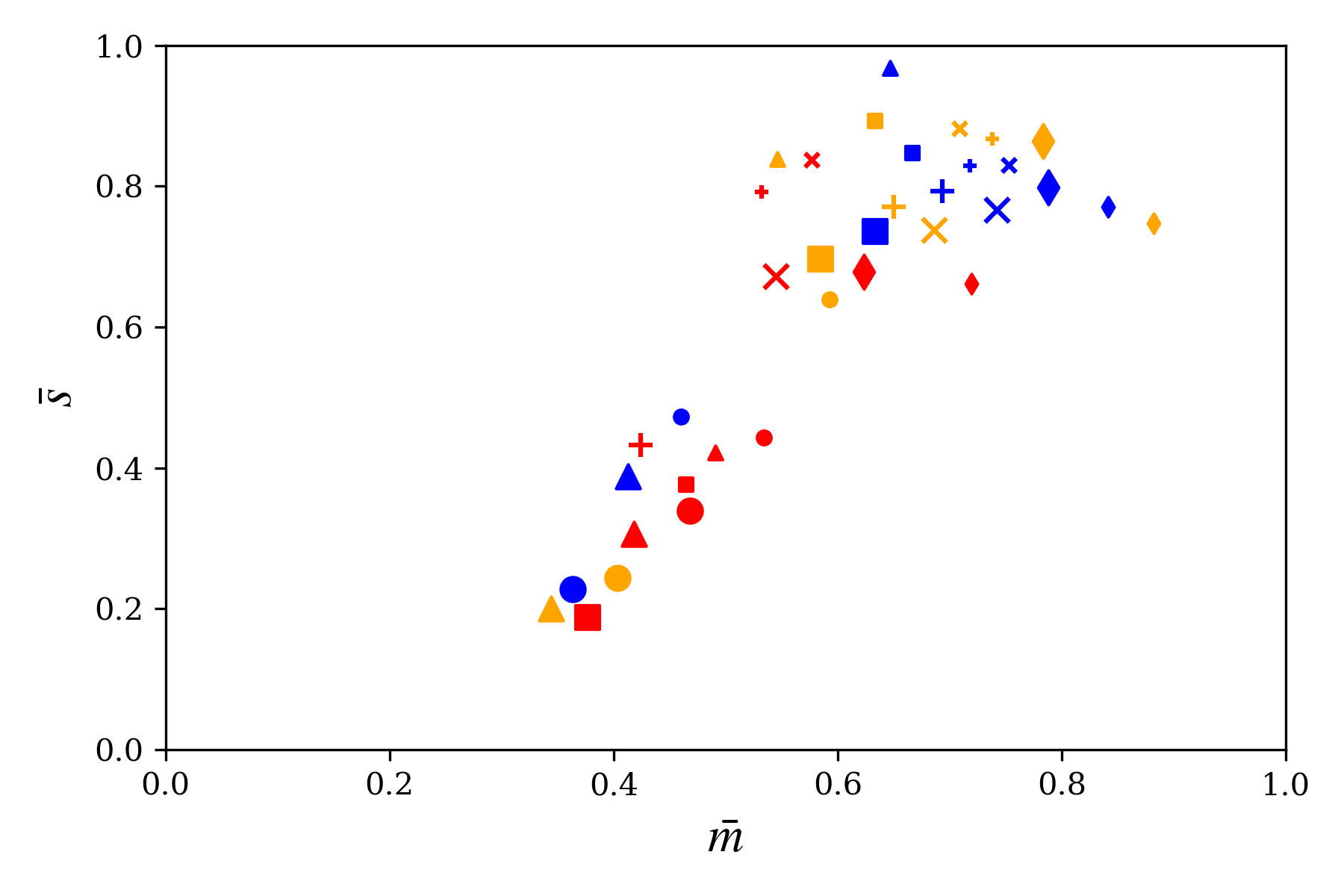}
\caption{The locations of star-system formation-site and final-position MSTs, on the ($\bar{m},\bar{s}$) plane. Magnetic field strengths are colour-coded: $B_{\rm o} = 0\, \mu$G (blue), $B_{\rm o} = 2\, \mu$G (yellow), and $B_{\rm o} = 5\, \mu$G (red). Filled circles, triangles, squares, pluses, crosses and diamonds correspond to collision velocities $\Delta u_{\rm o} = 2.4$, 2.8, 3.2, 3.6, 4.0 and $4.4$\, km\,s$^{-1}$, respectively. Smaller symbols indicate MSTs based on formation positions. Larger symbols indicate MSTs based on final positions at $t_{10\%}$.}
\label{fig:MoverS}
\end{figure}

The spatial distribution of the star-systems can be characterised by two parameters. The first is the normalised mean MST edge-length,
\begin{eqnarray}\label{EQN:mbar}
\bar{m}&=&\frac{{\cal N}- 1}{(A{\cal N})^{1/2}} \sum\limits_{n=1}^{n={\cal N}-1}\left\{\vphantom{y^a_b}m_n\right\}\,.
\end{eqnarray}
Here ${\cal N}$ is the number of star-systems, $A$ is the area of the convex hull enclosing their projected positions ($A$ is computed using \texttt{scipy.spatial.ConvexHull}), and $m_n$ is the length of the $n^{\rm th}$ edge on the MST. There are ${\cal N}\!-\!1$ edges on the MST.

The second is the normalised mean correlation length,
\begin{eqnarray}\label{EQN:sbar}
\bar{s}&=&\left(\frac{\pi}{A}\right)^{1/2} \frac{2}{{\cal N}({\cal N}-1)} \sum\limits_{i=1}^{i={\cal N}-1} \sum\limits_{j=i+1}^{j={\cal N}} \left\{\vphantom{y^a_b}s_{ij}\right\}\,.
\end{eqnarray}
Here $s_{ij}$ is the projected separation between star-systems $i$ and $j$, and the sum is now over all pairs of star-systems. Thus there are ${\cal N}({\cal N}-1)/2$ correlation lengths.

We note that the normalisations in Equations \ref{EQN:mbar} and \ref{EQN:sbar} means that the values of $\bar{m}$ and $\bar{s}$ do not reflect the overall physical size of the star-system distribution, but just the sizes of individual star-system separations relative to the overall extent of the distribution.\footnote{The values of $\bar{m}$ and $\bar{s}$ can be combined to compute $\mathcal{Q}=\bar{m}/\bar{s}$ \citep{CartwrightAWhitworthA2004MN348p589}. ${\cal Q}$ can then be used (i) to distinguish fractal star clusters with no central concentration, from spherical star clusters with a smooth radial density gradient, and hence (ii) to obtain either a measure of the fractal dimension, ${\cal D}$, or the radial density exponent, $-d\!\ln(n_\star)/d\!\ln(R)$. However this interpretation is not pertinent here. The clusters are neither smooth, nor spherical, nor fractal (in the sense of having an approximately self-similar nested hierarchy of structures).}

Figure \ref{fig:MoverS} shows the combinations of $\bar{m}$ and $\bar{s}$ for the MSTs of the initial and the final star-system locations. The $[\bar{m},\bar{s}]$ combinations fall into two groupings, a {\it lower grouping} at $\bar{m}\!\lesssim\!0.5$ and $\bar{s}\!\lesssim\!0.5$, representing \HF\ morphologies; and an {\it upper grouping} at $\bar{m}\!\gtrsim\!0.5$ and $\bar{s}\!\gtrsim\!0.7$, representing \SPW\ morphologies. In almost all cases the shift from the formation $[\bar{m},\bar{s}]$ combination to the corresponding final $[\bar{m},\bar{s}]$ combination is towards lower $\bar{m}$ and lower $\bar{s}$, as individual star-systems tend to migrate closer to their nearest neighbours as time advances.

For clear \HF\ morphologies (strong $B_{\rm o}$ and/or small $\Delta u_{\rm o}$) the formation and final $[\bar{m},\bar{s}]$ combinations are both in the {\it lower grouping} because most -- but not all -- star-systems are formed in close proximity to one another in the central hub. A few are formed in the filaments feeding the central hub, and therefore the convex hull has an area much larger than the hub.

For clear \SPW\ morphologies (weak/no $B_{\rm o}$ and/or large $\Delta u_{\rm o}$) the formation and final $[\bar{m},\bar{s}]$ combinations are both in the {\it upper grouping} because most -- but not all -- star-systems are formed in isolation, from a single core. The numerical experiments do not have enough resolution to capture the formation of a small sub-cluster from each core, although this is what we would expect to happen in nature \citep[see e.g. ][]{AmbroseHWhitworthA2024MN535p3700, AmbroseHWhitworthA2025MN541p3728}.

For transitional setups in the strip of ($B_{\rm o},\,\Delta u_{\rm o}$) parameter space intermediate between clear \HF\ morphologies and clear \SPW\ morphologies, the formation $[\bar{m},\bar{s}]$ combinations are in the {\it upper grouping} mimicking the dispersed formation sites of a \SPW\ morphology, and the final $[\bar{m},\bar{s}]$ combinations are in the {\it lower grouping} mimicking the centrally concentrated final sites of an \HF\ morphology.

\section{Discussion}\label{SEC:Discussion}

\subsection{Overview}

The numerical experiments presented here pertain to artificially simplified setups. This approach has been adopted so that the parameter space defining the initial conditions is small, and therefore it is more straightforward to identify cause and effect. In this case there are just two parameters, the initial magnetic field, $B_{\rm o}$, and the collision velocity, $\Delta u_{\rm o}$.

Equivalently we could use the dimensionless parameters $\beta_{\rm o}$ and ${\cal M}_{\rm o}$. Here, $\beta_{\rm o} =B_{\rm o}/B_{\rm crit}$ (where $B_{\rm crit}=2\pi G^{1/2}\rho_{\rm o}R_{\rm o}$ is the field strength required for magnetic support to balance the self-gravity of the cloud), and ${\cal M}_{\rm o}=u_{\rm o}/a_{\rm o}$ is the Mach Number of the cloud velocities.

It turns out that, in effect, the results form a one-parameter family, with an increase in the initial magnetic field strength being equivalent to a reduction in the collision velocity. The single parameter is $\Delta u_{\rm o}\{1-(B_{\rm o}/B_{\rm crit})^2\}\;$ or $\;{\cal M}_{\rm o}\{1-\beta_{\rm o}^2\}$.

\subsection{Limitations of the model}

In this section we discuss the additional physical effects that would be most likely to change our results, and how they might change them.

\subsubsection{'Turbulent' clouds}

Real clouds have internal structure, manifest as both systematic and random variations in magnetic field, radiation field, density, temperature, chemical composition, and bulk velocity. Including these variations would disrupt the regularity and coherence of the shock-compressed layer.

For setups with low (or no) initial magnetic field, and/or high collision velocity, there should still be widely distributed cores, as in the \SPW\ morphology, but there will be a dispersion in the times at which these cores start forming stars, and therefore also a dispersion in the times at which they finish forming stars.

For setups with high initial magnetic field and/or low collision velocity, there should still be an \HF\ morphology, but there will be a dispersion in the times at which filaments form. As filaments are dragged into radial orientations, smaller filaments and those that attempt to form later, will be assimilated by larger filaments and those that have formed earlier. The effect will be to reduce the number of dominant large-scale filaments, thereby producing more realistic \HF\ morphologies.

\subsubsection{Collisions at finite impact parameter}

Real clouds do not collide head-on. Collisions at finite impact parameter deliver systematic angular momentum, ${\boldsymbol L}_{\rm o}$, to the shock-compressed layer, and the resulting shear inhibits contraction orthogonal to ${\boldsymbol L}_{\rm o}$. As a result, the dominant filaments should align preferentially orthogonal to ${\boldsymbol L}_{\rm o}$. The filaments in observed \HF\ morphologies often show a preferential orientation -- rather than being statistically isotropic as in the head-on \CCCs\ modelled here.

\subsubsection{Alignment of the magnetic field with the collision axis}

There should be fundamental differences in the phenomenology of \CCCs\ if the magnetic field is not aligned with the collision velocity.

In the extreme case where the field is orthogonal to the collision velocity, the gas will only be significantly compressed if the initial field is much smaller, $B_{\rm o}\ll5\,\upmu{\rm G}$, and/or the collision velocity is much higher, $\Delta u_{\rm o}\gg 5\,{\rm km\,s^{-1}}$, than the values treated in the setups considered here.

If the field is at an intermediate angle, $\theta_{uB},$ to the collision velocity (intermediate between $0^\circ$ and $90^{\circ}$), the constraints on $B_{\rm o}$ and $\Delta u_{\rm o}$ for significant compression are relaxed (relative to the orthogonal case), but they are still quite restrictive for all but the smallest $\theta_{uB}$.

It seems likely that the velocities of clouds involved in those \CCCs\ that initiate star formation are preferentially aligned with the local magnetic field.

\subsubsection{Mass resolution}

The mass functions of star-systems might change somewhat if the resolution were improved, and this needs to be explored. In particular, a core formed in the shock compressed layer would be expected to spawn a small sub-cluster, and the sub-cluster would likely include some stars that are too small to have been resolved here.

Since (a) each SPH particle has mass $m_{\mbox{\tiny SPH}}=0.001\,{\rm M}_{_\odot}$ and $\sim 57$ neighbours, these experiments cannot properly resolve star-systems below the Hydrogen Burning Limit at $\sim 0.075\,{\rm M}_{_\odot}$ (i.e. Brown Dwarfs and Planetary Mass Objects), and therefore the mass functions are truncated artificially below this mass.

However, it is also the case that, in the Milky Way near the Sun, stars below the Hydrogen Burning Limit contribute very little to the overall mass budget (less than $2\%$), so the mass functions obtained here may be an acceptable approximation to star-system mass functions.

\subsubsection{Other limitations, and termination of the experiments}

We should not expect fundamental differences in the large-scale phenomenology of star formation if the two colliding clouds have different masses and/or radii, provided the clouds have comparable surface-densities.

At the volume- and surface-densities in the experiments, the sound-speed of the gas is not expected to vary much (typically between $\sim 0.15$ and $\sim 0.30\,{\rm km\,s^{-1}}$), and the degree of ionisation is unlikely to fall so low that the field and the gas decouple. Therefore we should not expect fundamental differences in the large-scale phenomenology of star formation initiated by \CCCs\ if the thermodynamics, chemistry and radiation transport are treated in more detail, or the assumption of ideal MHD is relaxed.

The neglect of feedback is a significant concern. Feedback is likely to be especially important in the monolithic star clusters that form more massive stars (Georgatos \& Whitworth, in prep.). It may also be important in the evolution on scales below our resolution limit ($\sim 10^3\,{\rm AU}$), where small sub-clusters and binary/multiple systems are formed. In real clouds star formation is self-limiting, and appears to terminate once 5 to 20\% of the available mass has been converted into star-systems. The artificial device of terminating the experiments once 10\% of the mass has been converted into star-systems is therefore an important limitation.

Since star formation is a chaotic process, we have performed three realisations for each $(B_{\rm o},\,\Delta u_{\rm o})$ combination, and collated the results to obtain the statistics presented in Figures \ref{fig:Width}, \ref{fig:Angles}, \ref{fig:Bscale}, \ref{fig:firstsink}, \ref{fig:Ridge}, \ref{fig:MoverS} and accompanying text. Whilst it would be better to have a larger number of realisations, the systematic trends with increasing $\Delta u_{\rm o}\{1-(B_{\rm o}/B_{\rm crit})^2\}$ that constitute the main conclusions of this paper are robust, i.e. more pronounced than any stochastic variance.

\subsection{ Comparison with observations}

 \CCCs and \HFSs\ appear to be important pathways for forming both low-mass stars \citep[e.g.][]{LorenRB1976ApJ209p466, MyersP2009ApJ700p1609, AndrePetal2014PPVI27} and high-mass stars \citep[e.g.][]{MotteFetal2018ARAA56p41, SeshadriA2024MN527p4244, MaityAetal2025AJ169A325}. In this section we review observational evidence for the involvement of \CCCs\ in triggering the star formation, the role of \HFSs\ in regulating the subsequent star formation, and how the numerical experiments reported here relate to this evidence.

\subsubsection{Cloud-Cloud Collision Velocities}

 Observed putative \CCCs\ usually involve higher collision velocities than those used in our experiments. For example, \citet{PerettoNetal2014AA561A83, WilliamsGetal2018AA613A11} and \citet{WangJWetal2022ApJ931A115} argue that the SDC13 \HFS\ is the result of a \CCC\ at $10\;{\rm to}\;20\,{\rm km\,s^{-1}}$;~ \citet{FukuiYetal2018aPASJ70S44, FukuiYetal2018bPASJ70S46, FukuiYetal2018cApJ859p166} argue that the star formation in -- respectively -- GM24, S116/S117/S118, and the Orion Nebula Cluster, has been triggered by \CCCs\ at velocities in the range $5\;{\rm to}\;10\,{\rm km\,s^{-1}}$.

\subsubsection{ Self-Organised Criticality}\label{SEC:GenConsid}

The relatively low collision velocities that we explore here are informed by a picture in which star formation is regulated by {\it Self-Organised Criticality}. In this picture, smaller clouds merge to form larger clouds, and conversely larger clouds break up into smaller clouds. This continuous restructuring (normally described as {\it turbulence}) proceeds until a cloud is formed somewhere -- presumably by the merger of smaller clouds -- with sufficiently high mass and density, sufficiently low temperature and velocity dispersion, and sufficient coherence, that it collapses and fragments to form a star or group of stars

 Within a molecular cloud complex, mergers occur at different times and on a range of scales. Therefore, even in a single cloud complex, there may be many more-or-less independent on-going mergers. Most of these mergers are the outcome of low-velocity collisions between the transient clouds that constitute the turbulent substructure within the larger cloud complex. These clouds need not have been long-lived coherent structures existing in isolation before the collision.

It follows that the distinction between star formation in a single cloud and star formation triggered by a \CCC\ is moot. A cloud, that on the large scale appears isolated and single, contains smaller sub-clouds that repeatedly bump into one another. The clouds in our numerical experiments represent these sub-clouds. It is for this reason that they have velocities drawn from Larson's scaling relations, as described in Section \ref{Sec:NumercialMethods}. They have uniform-density and spherical boundaries purely so that we can define the initial conditions for a collision using a small number of parameters -- not because we believe this to be a good representation of reality. For the same reason, they only collide head-on.

\subsubsection{ Experiments, not Simulations}

As stated in Section \ref{SEC:Intro}, the \CCCs\ that we model are {\it numerical experiments} rather than {\it simulations}. Our aim is to explore the phenomenology of low-velocity \CCCs\ and the overall patterns of star formation, in the spirit of `proof-of-concept', rather than to reproduce realistic star-formation regions that might be compared with observation. In order to make such a comparison, we would need (i) to consider much more chaotic initial conditions (reflecting a range of cloud masses, sizes, turbulence levels, etc.), (ii) to collide them with a range of mass-ratios, velocities and impact parameters, and (iii) to do multiple realisations for each parameter set -- since the processes involved in star formation are highly non-linear and chaotic. Such a study lies outside the scope of this paper.

\subsubsection{Evidence for Cloud-Cloud Collisions}

As stressed in the Introduction, robust observational evidence for \CCCs\ playing a critical role in triggering star formation is hard to obtain, for three main reasons. Firstly, unambiguous signatures of ongoing or recent star formation tend to be seen after the collision is spent.

Secondly, the collision velocity can only be measured if it is sufficiently close to the line of sight, which means that there is then confusion between the two clouds, and it is hard to demonstrate conclusively that the two clouds are actually in contact \citep{PriestleyFWhitworthA2021MN506p775}.

Thirdly, the shock-compressed layers produced by \CCCs\ are hard to identify observationally. Seen close to face-on their thinness along the line of sight is masked unless there is an accurate estimate of the volume-density. Seen close to edge-on they are indistinguishable from filaments.

The most widely used procedure for detecting a \CCC\ is to look for a bridging feature in the PV map of a molecular-gas tracer like CO \citep{HaworthTetal2015aMN450p10, HaworthTetal2015bMN454p1634}. This is the technique used, for example, by \citet{FukuiYetal2018cApJ859p166} to infer that the ONC was formed by a \CCC. ~As pointed out by \citet{PriestleyFWhitworthA2021MN506p775}, it may be necessary to also observe a denser gas tracer like NH$_3$ or HCN, in order to confirm that the two clouds are actually colliding, and not just superposed by chance along the same line of sight.

\subsubsection{Hub-filament systems}

Hub filament systems are ubiquitous, and cover a range of scales. However their signature is essentially topological, and there are no robust statistical metrics to support a distinctive taxonomy, or facilitate detailed comparison between observation and theory. Statistical metrics are needed because star formation is chaotic, and therefore no two \HFSs\ are the same in detail. The following properties might, in the future, provide fruitful metrics: mass, linear extent, number and orientation of filaments \citep[see][]{PerettoNetal2022EPJWC257p00037}, velocity field, structure and strength of magnetic field, evolutionary stage, and stellar mass function. However, at present most of these parameters are hard to estimate, and even harder to estimate in a large sample. As emphasised by \citet{DewanganLetal2026AJ171p69}, each system is unique, and there is a limited number of well observed systems.

\subsubsection{The masses and extents of known Hub-Filament Systems}

{Well-observed \HFSs\ tend to be more massive and extensive than those formed in our experiments. For the large sample analysed by \citet{KumarMetal2020AA642A87}, and for specific cases treated in greater detail, for example, SDC355 \citep{PerettoNetal2013AA555A112}, SDC13 \citep{PerettoNetal2014AA561A83, WilliamsGetal2018AA613A11, WangJWetal2022ApJ931A115}, Musca \citep{BonneLetal2020aAA641A17, BonneLetal2020bAA644A27}, Mon R2 \citep{KumarMetal2022AA658A114, DewanganLetal2025AJ169p80}, the G11 IRDC \citep{DewanganLetal2024MNRAS527p5895}, W3(OH), W3 Main, S106 \citep{KumarMetal2025AA703A74}, W33 \citep{ZhouJWetal2023MN519p2391, DewanganLetal2026AJ171p69}, the extents range from $\sim 2\,{\rm pc}$ to $\sim 20\,{\rm pc}$. In contrast, the \HFSs\ formed in our numerical experiments have extents $\lesssim 0.5\,{\rm pc}$. This difference simply reflects a selection effect. Larger \HFSs\ are easier to detect and easier to observe, and therefore they dominate the observed sample. Conversely, by modelling smaller \HFSs\ our experiments are able to resolve forming star-systems with masses down to the Hydrogen Burning Limit. The physics involved should be essentially scale free between the large observed systems and our smaller modelled systems.

 The star-forming complex W33 \citep{ZhouJWetal2023MN519p2391, DewanganLetal2026AJ171p69} provides a good example of the connection between \HFSs\ and star formation. Multiple \HFSs\ are identified within the cloud, each containing a rich embedded protostellar population. The hubs coincide with column-density peaks and enhanced velocity dispersion, but most lack signatures of ionising feedback, indicating a relatively early evolutionary stage. The median extent of the \HFSs\ in W33 is $\sim$2.4\,pc.

Comparable morphologies are widely reported. The infrared dark cloud SDC 335 \citep{PerettoNetal2013AA555A112} consists of six filaments feeding a hub containing two extremely massive cores; the size of this \HFS\ is $\sim$2\,pc. The G33 \HFS\ analysed by \cite{WangJWetal2020ApJ905A158} exhibits four major filaments converging onto a central hub. The system extends over $\sim 2$\, pc and shows evidence of massive protostars associated with the hub. JWST and ALMA observations of the G11P1 \HFS\ \citep{BhadariNetal2025AA694L18} reveal clear inflow signatures towards a massive hub that hosts a protostar; the extent of this particular \HFS\ is $\sim 1\,{\rm pc}$.

These observed structures are similar to the \HFSs\ formed in our experiments. In our low-velocity and/or high-magnetic-field experiments, the shock-compressed layer fragments into an \HFS\ with a characteristic extent of $\lesssim$0.5\,pc. As the collision velocity is decreased, or the initial magnetic field is increased, the number of filaments feeding the hub, and the overall extent of the system, both decrease.

\subsubsection{ Filament {\sc fwhm}s}

On the basis of dust emission, \citet{ArzoumanianDetal2011AA529L6, ArzoumanianDetal2019AA621A42} find that filaments in nearby star formation regions (a) tend to have a characteristic {\sc fwhm} of $\sim 0.1\,{\rm pc}$, and (b) tend to contain most of the dense star-forming gas and prestellar cores. A very similar distribution of filament widths is found in numerical simulations \citep[e.g.][]{PriestleyFWhitworthA2022aMN509p1494, PriestleyFWhitworthA2022bMN512p1407}. However, this universal filament {\sc fwhm} is not seen when looking at molecular-gas tracers \citep{PanopoulouGetal2017MN466p2529, PanopoulouGetal2022AA657L13}. In general, the distribution of widths found using gas tracers is systematically higher and broader. Moreover, filament {\sc fwhm}s are hard to estimate beyond a few hundred parsecs \citep{JuvelaMetal2012AA544A141}

The distribution of filament widths found in the experiments reported here lies between those reported by \citet{ArzoumanianDetal2019AA621A42} and \citet{PanopoulouGetal2017MN466p2529}. It is closer to \citet{PanopoulouGetal2017MN466p2529} (i.e. broader filaments) when the magnetic field is stronger. The filaments in the more massive, more extended observed \HFSs\ are also broader \citep[e.g.][]{WilliamsGetal2018AA613A11, WangJWetal2022ApJ931A115}, suggesting that the magnetic field is strong.

Parenthetically we do not see any evidence for fibre bundles \citep{HacarAetal2013AA554A55, TafallaMHacarA2015AA574A104, HacarAetal2017AA606A123, HacarAetal2018AA610A77}. With more realistic initial conditions (specifically turbulent substructure and a non-uniform magnetic field in the initial clouds), accretion onto a filament would deliver parcels of gas (a) at a less regular rate, and (b) with different flux-to-mass ratios. This would result in coherent radial oscillations involving different elements of the filament, and these might dissipate quite slowly given the elasticity inherent in their magnetic radial support, thereby producing fibres.

\subsubsection{The coherence of Hub-Filament Systems}

\citet{PerettoNetal2022EPJWC257p00037} note that {\it ``there is no clear definition of what a hub is''}, and introduce a {\it convergence parameter} which reflects the number of filaments anchored in a hub, and the extent to which the filaments point towards the hub. Using this metric they show that very few filaments are associated with hubs, but most high-mass star formation occurs in hubs. The \HFSs\ formed in our experiments have high convergence parameters (i.e. a large number of essentially radial filaments). However, this is artificial. It is the result of our very simplistic initial conditions, i.e. spherical, uniform-density clouds that collide head-on.

\subsubsection{Evidence for longitudinal filament fragmentation}

In the numerical experiments, filament fragmentation tends to occur initially at the intersections of filaments. Consequently the spacing of filaments, and the masses of the cores that condense out at their intersections, are lower at higher collision velocity and/or lower initial magnetic field. In contrast, at lower collision velocity and/or higher initial magnetic field, the filaments are dragged into radial orientations and there are fewer intersections; the fragmentation wavelength along the length of a filament, and the masses of the resulting cores, both then tend to be larger. In the experiments, the separations between neighbouring filament spines, and between neighbouring cores along a filament, typically range from $\sim 0.1\,{\rm pc}$ to $\sim 0.3\,{\rm pc}$. These values are comparable with observed estimates.

\subsubsection{The relationship between filament spines and the local magnetic field}

\citet{ArzoumanianDetal2021AA647A78} explore the alignment between the spines of filaments and the local magnetic field in the NGC 6334 \HFS. Broadly speaking the magnetic field is poorly aligned with the spines of faint outlying filaments with low column-densities; and the alignment steadily increases in more pronounced filaments at higher column-densities, closer to the hub. Similar results have been found by other authors \citep[e.g.][]{KumarMetal2025AA703A74}, and this is essentially what we find in our numerical experiments (see Figures \ref{fig:Lines} and \ref{fig:Angles}, and the accompanying discussion in Section \ref{SEC:Alignment}). At low-density the gas tends to flow along the local magnetic field and onto a filament; consequently there is no tendency for the field and the filament spine to be aligned with one another. Once the gas accretes onto a filament its density increases, and it then flows along the filament carrying the field with it, so the field and the filament spines are approximately parallel to one another at higher densities.

\subsubsection{The magnetic field strength}

In our numerical experiments, the magnetic field in the filaments is typically between $100\,\upmu{\rm G}$ and $1000\,\upmu{\rm G}$. significantly higher than in the initial clouds ($\lesssim 5\,\upmu{\rm G}$). These values are comparable with the estimated values in \HFSs\ \citep[e.g.][]{HwangJetal2022ApJ941A41}.

\subsubsection{The spatial distribution of stars formed in a hub}

The stars formed in a hub are initially densely packed. There is a large supply of gas into the hub, which is why a few stars there grow to high mass by competitive accretion. The velocity dispersion in the hub is typically in the range $1\;{\rm to}\;4\,{\rm km\,s^{-1}}$. A few low-mass stars are ejected dynamically.

\subsubsection{The stellar mass distribution}

The mass distribution for the star-systems produced in the experiments depends systematically on the collision velocity and the magnetic field strength, as shown on Figure \ref{fig:Ridge}. Reducing the collision velocity has a very similar effect to increasing the magnetic field, in the sense that both effects slow down the build-up and fragmentation of the shock-compressed layer to produce filaments. This gives more time for the filaments to be dragged into radial orientations by \LAT.

Thus, as the collision velocity is reduced, and/or the initial magnetic field is increased, the mass function first shifts to higher masses. As more and more mass falls into the central hub, a few star-systems can grow to large values by competitive accretion. At the same time a low-mass tail grows due to dynamical ejections.

Then as the collision velocity is reduced further and/or the magnetic field is increased further, the material falls into the hub {\it so slowly} that most of it has been converted into star-systems before it arrives in the hub. Consequently there is little opportunity for competitive accretion and few, if any, very high-mass stars are formed. At the same time, due to the extra magnetic support, the masses of the low-mass stars formed by filament fragmentation systematically increase.

\subsubsection{Mass segregation in hubs}

Star formation in \HFSs\ naturally produces mass segregation, and a delay between low-mass star formation (which has already started in the filaments, before they have fed much gas into the hub, and then continues in the hub) and high-mass star formation (which is restricted to the high-densities in the hub), as argued by \citet{FukuiYetal2018cApJ859p166} and \citet{KumarMetal2020AA642A87}. The same feature is seen in our experiments, with high-mass stars forming only once enough gas has accumulated in the hub. However, the delay between low- and high-mass star formation in our experiments is only $\sim0.1\,{\rm Myr}$, whereas \citet{FukuiYetal2018cApJ859p166} estimate a delay of $\sim1\,{\rm Myr}$ in the Orion Nebula Cluster. This difference is partly -- but only partly -- attributable to the smaller size of the clouds involved in our numerical experiments, and the lack of turbulence, both of which accelerate star formation in the numerical experiments.

\subsection{ Alternative mechanisms for forming Hub-Filament Systems}

\citet{MyersP2009ApJ700p1609} proposes that \HFSs\ form following the collapse of a cloud to a pancake, followed by fragmentation of the more diffuse outer layers of the pancake into filaments. This scenario is closely related to the one proposed here, except that the velocity of the gas falling onto the pancake is generated by the self-gravity of the cloud and external compression, whereas the velocity of the gas flowing into the shock-compressed layer in our experiments is generated by self-gravity and the bulk velocities of the colliding clouds. The velocities in the numerical experiments tend to be somewhat higher, and to deliver greater ram pressure.

\citet{KumarMetal2020AA642A87} develop a scenario for forming \HFSs\ in which two independent filaments collide, and this then enhances the density locally at their intersection. Gas then converges towards this density enhancement, forming a hub \citep[e.g.][]{PanjaAetal2023ApJ958A17}. Low-mass star formation proceeds slowly in the filaments, whilst high-mass star formation is concentrated in the hub \citep[see the cartoon in Figure 14 of][]{KumarMetal2020AA642A87}. This scenario assumes the pre-existence of coherent filaments moving independently of one another, and typically produces \HFSs\ with three or four filaments. It requires that the pre-existing filaments are relatively stable against fragmentation and star formation before they collide; and that their collision to produce a hub (which in the first instance can only approximately double the line-density locally) increases the gravitational pull of the hub sufficiently to drag in the sections of filament on either side and initiate massive star formation in the hub. In this scenario the magnetic field in a filament should not be amplified by compression until it converges on the central hub, and therefore it should be lower, ~$\lesssim 10\,\upmu{\rm G}$,~ than in our experiments, where it reaches values ~$\gtrsim 100\,\upmu{\rm G}$.

\section{Conclusions}\label{SEC:Conclusions}

We have investigated the role of magnetic fields in head-on \CCCs\ between spherical clouds with mass $500\,{\rm M}_{_\odot}$ and radius $2\,{\rm pc}$. The clouds have uniform density and the magnetic field is aligned with the collision velocity. Therefore different setups are characterised by the initial field strength, $B_{\rm o}$, and the collision velocity, $\Delta u_{\rm o}$.

\begin{itemize}

\item Although we have considered setups defined by two independent parameters, $B_{\rm o}$ and $\Delta u_{\rm o}$, most of the results appear to subscribe to a one-dimensional family.  An increase in the magnetic field is equivalent to a decrease in the collision velocity, in that both have the effect of slowing down the process of \ACC, and thereby creating more time for \LAT\ of the shock compressed layer. This in turn shifts the pattern of star formation from the \SPW\ morphology to the \HF\ morphology, hence away from dispersed star formation in small relatively isolated cores with a relatively narrow mass function, and towards star formation in a monolithic centrally concentrated star cluster with a broader mass function.

\item The switch from the \SPW\ morphology for faster `magnetic-weak' setups, to the \HF\ morphology for slower `magnetic-strong' setups, occurs at collision velocity
\begin{eqnarray}
\Delta u_{\rm o}\;\;\sim\;\;3\,{\rm km\,s^{-1}}\left\{1\,-\,(B_{\rm o}/10\upmu{\rm G})^2\right\}^{-1}.
\end{eqnarray}

\item For slower and/or `magnetically-strong' setups, $\Delta u_{\rm o}\lesssim 2.8\,{\rm km\,s^{-1}}\,\{1-(B_{\rm o}/10\upmu{\rm G})^2\}^{-1}$, the shock-compressed layer fragments into an \HF\ morphology of radially oriented filaments that feed material into a central hub, and a few star-systems condense out of the filaments and fall into the hub alongside the gas. Once in the hub more star-systems condense out and compete to accrete the remaining gas. The filaments tend to be somewhat broader, $0.04\,{\rm pc}\lesssim${\sc fwhm}$\lesssim0.17\,{\rm pc}$. A small number of star-systems grow to high mass by virtue of being able to compete for additional mass from the large reservoir in the central monolithic hub. The stars, in particular the high-mass stars, are strongly concentrated within the central hub. The magnetic field in the filaments reaches quite high values, $\sim 100\,\upmu{\rm G}$ to $\sim 1000\,\upmu{\rm G}$.

\item For faster and/or `magnetic-weak' setups, $\Delta u_{\rm o}\gtrsim 3.2\,{\rm km\,s^{-1}}\,\{1-(B_{\rm o}/10\upmu{\rm G})^2\}^{-1}$, the shock-compressed layer fragments into a \SPW\ morphology of filaments, relatively small and isolated prestellar cores condense out, often at the intersections of the filaments, and these cores then spawn a dispersed array of star-systems. The filaments tend to be quite narrow, $0.02\,{\rm pc}\lesssim${\sc fwhm}$\lesssim0.10\,{\rm pc}$. High-mass star-systems, and therefore presumably high-mass stars ($\gtrsim 10\,{\rm M}_{_\odot}$) are rare, because the individual cores have small mass reservoirs. The stars are quite widely distributed, and there are few high-mass ones. The magnetic field in the filaments is relatively weak, $\sim 10\,\upmu{\rm G}$ to $\sim 100\,\upmu{\rm G}$.

\item There are also transitional setups for intermediate values of $\Delta u_{\rm o}\sim 3\,{\rm km\,s^{-1}}\,\{1-(B_{\rm o}/10\upmu{\rm G})^2\}^{-1}$, and a clear monotonic sequence in the statistical descriptors of star formation --- filament widths, star-system mass functions, initial and final spatial distributions of star-systems, degrees of magnetic field amplification --- with increasing $\Delta u_{\rm o}\{1-(B_{\rm o}/10\upmu{\rm G})^2\}$.

\item The magnetic field in a filament tends to be aligned with the spine of the filament, but there are also places where the field is approximately perpendicular to the spine and the field reverses direction across the filament.

\item Once star formation starts, the rate of star formation is essentially independent of the initial setup, and of order $10^3\,{\rm M_{_\odot}\,Myr^{-1}}$.

\end{itemize}

\section*{Acknowledgments}

We thank the referee for their constructive comments on the original draft of this paper, which have lead to significant improvements.  TG gratefully acknowledges the receipt of an STFC PhD studentship. The numerical experiments were all performed on the Cardiff University Advanced Research Computing facility (ARCCA).\\\\

\section*{Data Availability}
 
The results of the experiments are available upon request to TG.

\bibliographystyle{mnras}
\bibliography{cleaned} 

\bsp	
\label{lastpage}
\end{document}